\documentclass[11pt,a4paper]{article}
\usepackage{graphicx}
\usepackage{jcappub}

\graphicspath{{./figs/}}

\newcommand{\simgt}{\lower.5ex\hbox{$\; \buildrel > \over \sim \;$}}
\newcommand{\simlt}{\lower.5ex\hbox{$\; \buildrel < \over \sim \;$}}

\makeatletter
\newsavebox{\@parc@ption}
\def\parcaption#1{%
\sbox{\@parc@ption}{\shortstack[l]{#1}}%
>\setbox\@tempboxa\hbox{\csname fnum@\@captype\endcsname}%
\@tempdima\columnwidth \advance\@tempdima-\wd\@tempboxa
\@tempdimb\@tempdima 
\ifdim\wd\@parc@ption>\@tempdimb \@tempdima\@tempdimb
\else\@tempdima\wd\@parc@ption\fi
\sbox{\@tempboxa}{\parbox[t]{\@tempdima}{#1}}%
\caption{\usebox{\@tempboxa}}}
\makeatother

\title{Testing chameleon gravity with the Coma cluster}

\author{Ayumu Terukina${}^{1}$,Lucas Lombriser${}^{2,3}$
Kazuhiro Yamamoto${}^{1,4}$, \\
David Bacon${}^{2}$, Kazuya Koyama${}^{2}$, Robert C. Nichol${}^{2}$
}
\date{\today}

\affiliation{
${}^{1}$Department of Physical Science, Hiroshima University,
Higashi-Hiroshima, \\Kagamiyama 1-3-1, 739-8526,~Japan\\
${}^{2}$Institute of Cosmology and Gravitation, University of Portsmouth, Dennis Sciama Building,
Portsmouth, PO1 3FX, UK\\
${}^{3}$Institute for Astronomy, University of Edinburgh, Royal Observatory,\\
Edinburgh, EH9~3HJ, UK\\
${}^{4}$Hiroshima Astrophysical Science Center, Hiroshima University, Higashi-Hiroshima, \\
Kagamiyama 1-3-1,  739-8526, Japan
}

\date{\today}

\abstract{
We propose a novel method to test the gravitational interactions in the outskirts of galaxy clusters.
When gravity is modified, this is typically accompanied by the introduction of an additional scalar degree of freedom, which mediates an attractive fifth force.
The presence of an extra gravitational coupling, however, is tightly constrained by local measurements.
In chameleon modifications of gravity, local tests can be evaded by employing a screening mechanism that suppresses the fifth force in dense environments.
While the chameleon field may be screened in the interior of the cluster, its outer region can still be affected by the extra force, introducing a deviation between the hydrostatic and lensing mass of the cluster.
Thus, the chameleon modification can be tested by combining the gas and lensing measurements of the cluster.
We demonstrate the operability of our method with the Coma cluster, for which both a lensing measurement and gas observations from the X-ray surface brightness, the X-ray temperature, and the Sunyaev-Zel'dovich effect are available.
Using the joint observational data set, we perform a Markov chain Monte Carlo analysis of the parameter space describing the different profiles in both the Newtonian and chameleon scenarios.
We report competitive constraints on the chameleon field amplitude and its coupling strength to matter.
In the case of $f(R)$ gravity, corresponding to a specific choice of the coupling, we find an upper bound on the background field amplitude of $|f_{R0}|<6\times 10^{-5}$, which is currently the tightest constraint on cosmological scales.
}

\arxivnumber{}

\begin{document}
\maketitle

\flushbottom

\section{Introduction}
\vspace{3mm}

Modifications of the theory of gravity can serve as an alternative approach to using dark energy models
to explain the cosmic acceleration of our Universe \cite{review}.
Any covariant modification of General Relativity introduces an additional degree of freedom.
The chameleon model modifies gravity by introducing a scalar field in addition to the tensor field,
which is non-minimally coupled with the matter components and gives rise to a fifth force that can
be of the same order as the standard gravitational force.
The scope for extra gravitational forces is, however, severely constrained by experiments in the Solar System.
The chameleon model employs a screening mechanism of the scalar field which depends on the local matter
density~\cite{Chameleonmechanism,Mota} and allows it to evade these constraints; 
however in this model cosmic acceleration must be driven by the contribution of a cosmological constant or dark energy rather than being a genuine modified gravity effect~\cite{WHK}.
When the curvature of space-time is small, gravity remains modified, which renders galaxy clusters a
useful regime for testing modified gravity models: while the interior of a cluster may be screened,
the chameleon mechanism may not completely screen the modifications of gravity in the outer region
of the cluster ~\cite{Ob1,dynamicalmass,Zhao11,Lombriser1,Lombriser2,Lam,Yano,NarikawaYamamoto,Terukina,Mota2,NariII}.
When the chameleon field is coupled with the gas component, the fifth force due to the
chameleon field affects the gas density profile of the galaxy cluster.
This causes an additional pressure gradient that balances the extra force, which leads to a more
compact gas distribution in the cluster.
This effect has been used in~\cite{Terukina} to compare the X-ray temperature profile predicted
by the chameleon model with measurements of the Hydra A cluster, yielding an upper bound on the
asymptotic scalar field value at large distances of $\phi_\infty<10^{-4}~M_{\rm Pl}$ for a coupling
constant between the chameleon field and matter of $\beta=1$.

The chameleon model parameter $\beta$ determines the strength of the fifth force when it is not
screened (see Sec.~\ref{sec:chameleon}).
The second chameleon parameter, $\phi_\infty$, controls the effectiveness of the screening mechanism,
describing the transition from the inner region of a cluster where gravity may be Newtonian to
the outer region where the fifth force contributes~\cite{Chameleongravity,Chameleongravity2}.
The critical radius, where the transition occurs, is determined by both $\phi_\infty$ and $\beta$
[see Eq.~(\ref{cp})].
In the absence of environmental effects, we may regard $\phi_\infty$ as the cosmological background
value of the chameleon field.
When $\beta=\sqrt{1/6}$
the chameleon model reduces to a $f(R)$ gravity model~\cite{Starobinski,HS,Tsujikawa} with the
scalar field potential determined by the choice of $f(R)$~\cite{Brax2008}, a nonlinear function
of the Ricci scalar $R$ that is added to the Einstein-Hilbert action.
In this case, the parameter $\phi_\infty$ is proportional to the parameter $|f_{R0}|$ of the $f(R)$
model, where $f_{R0}$ is the present background value of the scalar field $d f(R)/d R$ (see
Sec.~\ref{sec:fRconstraints} for details).

In the presence of a chameleon force, due to its effect on the gas distribution, the hydrostatic
mass of a cluster if interpreted assuming Newtonian gravity will deviate from its underlying dark
matter distribution, which can be measured via weak gravitational lensing, resulting in different
mass estimates for the cluster
(see~\cite{Arnold} for a recent analysis of this mass bias in hydrodynamic simulations of $f(R)$
gravity).
Therefore, the combination of the gas and lensing measurements of a cluster may yield a powerful
probe of gravity if they give statistically different mass estimates, which are not due to other
astrophysical reasons.

\begin{figure}[t]
\begin{center}
  \includegraphics[width=0.6\hsize,clip]{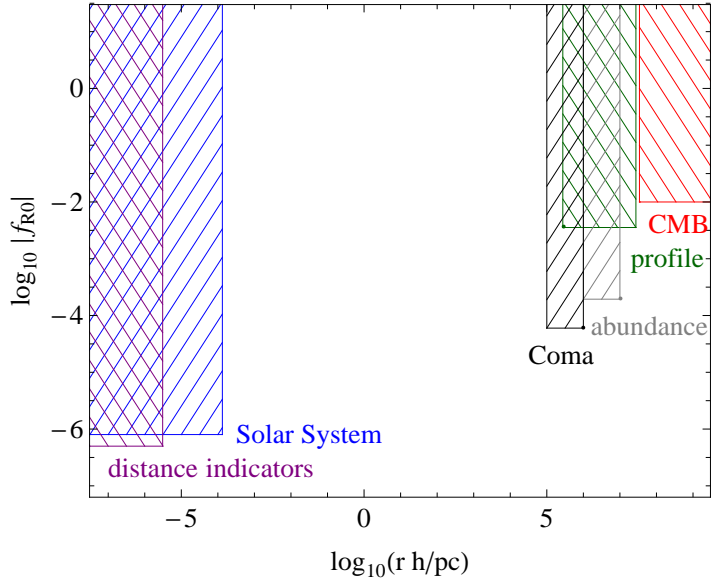}
\caption{
Comparison of our Coma cluster constraint to current constraints on $f(R)$ gravity from the Solar System~\cite{HS,Lombriser5}, distance indicators in unscreened dwarf galaxies~\cite{jain12}, the cosmic microwave background (CMB)~\cite{fRplanck1,fRplanck2}, cluster density profiles~\cite{Lombriser1} and abundance~\cite{Ob1,Lombriser3}.
The figure is adapted from~\cite{Lombriser1}.
Also compare to Fig.~2 (resp.~3) of~\cite{jain13,huterer13} for prospective constraints on $f(R)$ gravity.
\label{fig:mg_limit}
}
\end{center}
\end{figure}

In this paper, we demonstrate the operability of this method with the Coma cluster; a massive
cluster at a distance of approximately 100~Mpc, whose properties are measured with several
independent methods.
The \emph{Planck} team has, for instance, reported a precise observation of the Sunyaev-Zel'dovich
(SZ)~\cite{SZ} effect~\cite{Coma_5}.
Moreover, the X-ray surface brightness and X-ray temperature have been measured
in~\cite{Coma_2,Coma_3,Coma_4}, and weak lensing (WL) observations have been conducted by~\cite{Coma_6,Coma_7}.
We use the combination of these measurements to place tight constraints on $\beta$ and $\phi_\infty$.
To illustrate the effectiveness of our approach, in Fig.~\ref{fig:mg_limit}, we compare our result
to current constraints from cosmological, astrophysical, and local tests in the well studied case
that the chameleon model reduces to $f(R)$ gravity.
Our Coma constraint is currently the tightest constraint on cosmological scales.

An important element of our method is the reconstruction of the gas distribution in a cluster of galaxies under the influence of the fifth force.
In previous work~\cite{Terukina}, the hydrostatic equilibrium of the gas components was assumed when modelling the gas distribution of the Hydra A cluster in chameleon gravity.
Hydrostatic equilibrium may, however, not always be realised because of turbulence and bulk motions of the gas caused by mergers with other clusters and groups of galaxies, as well as infalling material.
The authors of~\cite{DSuto} have demonstrated that the cluster masses in numerical simulations, estimated under the assumption of hydrostatic equilibrium, can deviate from the true mass by up to 30\%, and that the deviation is explained by the acceleration term in the Euler equation.
We therefore carefully examine the systematics that deviations from the hydrostatic equilibrium in the Coma cluster may introduce on our results.

We first reconstruct the 3-dimensional profiles of the gas density, temperature, and pressure from the observational results using Newtonian gravity.
We then compare the mass estimates from the gas observations with the mass estimate from lensing, finding good agreement between them and that the assumption of hydrostatic equilibrium is a good approximation given the observational errors of the lensing mass.
Moreover, these mass estimates are only marginally affected by the inclusion of an extremised non-thermal pressure component, which has been calibrated to hydrodynamical simulations.

While non-thermal pressure and other deviations from the hydrostatic equilibrium enhance the hydrostatic mass estimate,
we find a strong decrease of the reconstructed hydrostatic mass when the chameleon fifth force is introduced.
The detection of an enhanced hydrostatic mass with respect to the lensing mass when interpreted in a Newtonian framework, may, therefore, be a smoking gun for modified gravity.
On the other hand, the effects of non-thermal pressure and chameleon force may become degenerate in the reconstruction, as the change in the hydrostatic mass by enhancing modifications of gravity can be compensated by increasing deviations from hydrostatic equilibrium.
Given the small effect of the non-thermal pressure compared to the effect from modifying gravity, however, we decide that it is safe to assume hydrostatic equilibrium of the gas, and perform our analysis under this assumption.

Finally, note that Fusco-Femiano, Lapi, and Cavaliere~\cite{Coma_1} recently investigated the consistency between the X-ray observations, from surface brightness and temperature, and the SZ measurement in the Coma cluster, adopting a ``Supermodel''.
The Supermodel expresses the profiles of density and temperature in the entropy-modulated equilibrium of the intracluster plasma within the potential wells provided by the dominant dark matter~\cite{CLFF09}.
This yields a direct link between the X-ray and the SZ observations based on the entropy profile.
They found a tension between the SZ and the X-ray pressure emitted by the plasma.
In our analysis, we confirm these results, by finding a similar tension between the SZ and the X-ray pressures.
However, the tension is mainly represented by the asymptotic difference of the
values of the pressure between the inner and the outer regions. On the other hand, the constraint on the
chameleon gravity model comes from the shape of the density profile in the intermediate regime, so we can nevertheless put a useful constraint on the chameleon model.

The paper is organised as follows: In Sec.~\ref{sec:2}, we review the hydrostatic equilibrium equations and hydrostatic mass, including a brief review of an analytic approximate solution of the scalar field profile in the cluster.
In Sec.~\ref{sec:3}, we perform a Markov chain Monte Carlo (MCMC) analysis to place constraints on the Newtonian and chameleon model parameter space.
We then discuss the systematic effects introduced by deviations from spherical symmetry, and study deviations from the hydrostatic equilibrium of the gas in Newtonian gravity by comparing the hydrostatic mass inferred from X-ray and SZ measurements with the lensing mass and analysing the effects of including non-thermal pressure in comparison to the effects from the chameleon force.
In Sec.~\ref{sec:conclusion}, we present our conclusions.
Finally, in Appendix~\ref{sec:3dprofiles}, we discuss our reconstruction method for the gas profiles.

\section{Hydrostatic and lensing mass in the presence of a chameleon force} \label{sec:2}

We describe the hydrostatic mass of a spherically symmetric system of gas and introduce the non-thermal pressure model, which we use to analyse deviations from hydrostatic equilibrium.
Then, we briefly review a derivation of an analytic approximate solution for the chameleon scalar field profile within a dark matter cluster, which we use to determine the effects on hydrostatic masses in the presence of the extra force.
Next, we compare the reconstructed hydrostatic masses, from different gas observations, with the observed lensing mass
and discuss the effect on the mass reconstruction when incorporating the non-thermal pressure model and the chameleon modification.

\subsection{Hydrostatic mass} \label{sec:hydrostaticmass}

We consider a spherically symmetric system of gas and dark matter.
In this case, we can write the equation for the gas component in hydrostatic equilibrium as
 \begin{eqnarray}\label{he}
  \frac{1}{\rho_{\rm gas}(r)}\frac{dP_{\rm total}(r)}{dr}  =-\frac{GM(<r)}{r^2},
 \end{eqnarray}
where $\rho_{\rm gas}$ is the gas density, $P_{\rm total}$ is the `total' gas pressure, including both thermal and non-thermal pressure,
and $M(<r)$ is the mass enclosed within the radius $r$.
This equation describes the balance between the gas pressure gradient and the gravitational force.
Note that we have not yet included the chameleon force (see Sec.~\ref{sec:chameleon}).
The total gas pressure can be written as the combination of the thermal pressure and the non-thermal pressure, $P_{\rm total}=P_{\rm thermal}+P_{\rm non-thermal}$.
Eq.~(\ref{he}) can then be rephrased as
 \begin{eqnarray}\label{HE1}
   M_{\rm}(<r)&=&M_{\rm thermal}(r)+M_{\rm non-thermal}(r)
 \end{eqnarray}
with the definitions:
 \begin{eqnarray}
  && M_{\rm thermal}(r)\equiv-\frac{r^2}{G\rho_{\rm gas}(r)}\frac{dP_{\rm thermal}(r)}{dr},
\label{mrrp}
\\
  && M_{\rm non-thermal}(r)\equiv-\frac{r^2}{G\rho_{\rm gas}(r)}\frac{dP_{\rm non-thermal}(r)}{dr}.
\end{eqnarray}
$M_{\rm non-thermal}$ is introduced to help mathematically describe the non-thermal pressure contribution to the total mass.
Note that $M_{\rm thermal}$ is expressed in terms of $P_{\rm thermal}$ and $\rho_{\rm gas}$
in Eq.~(\ref{mrrp}).
If we introduce the equation of state of gas, $P_{\rm thermal}=kn_{\rm gas}T_{\rm gas}$,
we can express the thermal mass in terms of $T_{\rm gas}$ and $\rho_{\rm gas}$ instead,
\begin{eqnarray}
&& M_{\rm thermal}(r)=-\frac{kT_{\rm gas}(r)r}{\mu m_{\rm p}G}
\left(\frac{d\ln \rho_{\rm gas}(r)}{d\ln r}+\frac{d\ln T_{\rm gas}(r)}{d\ln r}\right),
\label{mrrt}
\end{eqnarray}
where we have used $\rho_{\rm gas}=\mu m_{\rm p}n_{\rm gas}$
with the mean molecular weight $\mu$ and the proton mass $m_{\rm p}$.
The mean molecular weight of the fully ionised gas is defined by
$\mu(n_{\rm e}+n_{\rm H}+n_{\rm He})m_{\rm p}= m_{\rm p}n_{\rm H}+4m_{\rm p}n_{\rm He}$
with $n_{\rm e}=n_{\rm H}+2n_{\rm He}$, where $n_{\rm e}$, $n_{\rm H}$ and $n_{\rm He}$ is
the number density of electron, hydrogen, and helium, respectively.
Adopting the mass fraction of hydrogen $n_{\rm H}/(n_{\rm H}+4n_{\rm He})=0.75$,
we have $\mu=0.59$.

We define the fraction of the total pressure attributed to the non-thermal contribution by
 \begin{eqnarray}
   &&{P_{\rm non-thermal}}(r) \equiv g(r) P_{\rm total}(r).
 \label{pgp}
 \end{eqnarray}
Hence, using $P_{\rm total}=g^{-1}P_{\rm non-thermal}=(1-g)^{-1}P_{\rm thermal}$,
we may write
 \begin{eqnarray}
   P_{\rm non-thermal}(r)=\frac{g(r)}{1-g(r)}n_{\rm gas}(r)kT_{\rm gas}(r).
 \end{eqnarray}
According to hydrodynamical simulations~\cite{Battaglia,Shaw}, the non-thermal contribution to the total pressure can be modelled with the expression
 \begin{eqnarray}
  \label{grg}
   &&g(r)=\alpha_{\rm nt}(1+z)^{\beta_{\rm nt}}
   \biggl({r\over r_{500}}\biggr)^{n_{\rm nt}}\biggl({M_{200}\over
     3\times10^{14}M_\odot}\biggr)^{n_{\rm M}},
 \end{eqnarray}
where $\alpha_{\rm nt}$, $\beta_{\rm nt}$, $n_{\rm nt}$, and $n_{\rm M}$ are constants.
For illustration, and for an estimation of the effects from neglecting the non-thermal contribution, we adopt the parameter values
$(\alpha_{\rm nt},\beta_{\rm nt},n_{\rm nt},n_{\rm M})$ $=$ $(0.3,0.5,0.8,0.2)$,
which are the best-fit values in \cite{Shaw} with the exception of $\alpha_{\rm nt}$.
The best-fit value of $\alpha_{\rm nt}$ is $0.18$, which is an averaged
value over 16 simulated clusters.
We set $\alpha_{\rm nt}=0.3$, which is the maximum value obtained in the 16 clusters~\cite{Shaw}, in order to study the effect of the non-thermal pressure contribution in the extremised case.

We refer to Appendix~\ref{sec:3dprofiles} for our approach to reconstruction of the 3-dimensional profiles of $\rho_{\rm gas}$, $T_{\rm gas}$, and $P_{\rm thermal}$ from the gas observations via the X-ray temperature, X-ray surface brightness, and SZ effect, which enables us to estimate $M_{\rm thermal}$.
Using Eqs.~(\ref{pgp}) and (\ref{grg}), we can then estimate the non-thermal contribution $M_{\rm non-thermal}$ employing the results from hydrodynamical simulations.

\subsection{Chameleon fifth force} \label{sec:chameleon}

We now consider the effect on the hydrostatically inferred cluster mass profile when introducing the chameleon force.
The field equation of the chameleon field $\phi$ in a quasi--static system is given by \cite{Chameleonmechanism}
 \begin{eqnarray}\label{FieldEq}
   \nabla^2\phi=V_{,\phi}+\frac{\beta}{M_{\rm Pl}}\rho,
 \end{eqnarray}
where $\rho$ is the matter density, $V$ is the scalar field potential, $\beta$ is the coupling between the scalar field and matter, taken to be constant here, and $M_{\rm Pl}$ is the Planck mass.
We shall assume that the potential is a monotonic function of the scalar field, $V=\Lambda^{4+n}/\phi^n$ with constant exponent $n$ (see, e.g.,~\cite{Chameleongravity2,Terukina}).
Note that the choice of the potential is not essential to our analysis because the scalar field in the cluster is not sensitive to the parameters of the potential, $\Lambda$ and $n$, as will be described below.
Our results will also be applicable to the Hu-Sawicki $f(R)$ model in Sec.~\ref{sec:fRconstraints}, where the potential can be written approximately as
$V=V_0-\Lambda^{4+n}/\phi^n$ with $-1<n<0$ and a constant $V_0$ that yields cosmic acceleration.
We also assume $\beta\phi\ll M_{\rm Pl}$, so that the coupling is not too strong and we can use the weak-field approximation.
The chameleon fifth force is written as
 \begin{eqnarray}
   F_{\phi}=-\frac{\beta}{M_{\rm Pl}}\nabla \phi.
 \end{eqnarray}
Note that we are considering a model where both matter components, i.e., baryonic and dark matter, are coupled to the chameleon field.
In this scenario, both matter components are subject to the gravitational force and the chameleon force $F_{\phi}$.
This is, for instance, the case in $f(R)$ gravity models and any other chameleon theory that can be formulated in the Jordan frame with a single metric.
It is possible to consider a model where the baryonic component does not couple to the chameleon field~\cite{LiBarrowI,LiBarrowII}.
Such a model would not be constrained by our method as it would not introduce a difference between the hydrostatic and lensing masses.
Hence, we do not consider this possibility in this paper, nor the possibility of introducing different coupling strengths for the different components.

We further assume that the dark matter component dominates over the baryonic contribution in the cluster and that the matter density of the cluster $\rho$ is well described by a Navarro-Frenk-White (NFW)~\cite{NFWProfile} profile
 \begin{eqnarray}
   \rho(r)=\frac{\rho_{\rm s}}{r/r_{\rm s}\left(1+r/r_{\rm s}\right)^2}, \label{nfwfit}
 \end{eqnarray}
where the characteristic density $\rho_{\rm s}$, and characteristic scale $r_{\rm s}$, are fitted parameters.
The mass of the dark matter within the radius $r$ is then given by
\begin{eqnarray}\label{NFWmass0}
  M(<r)=4\pi \int_0^r dr r^2\rho(r) =4\pi \rho_{\rm s}r_{\rm s}^3 \left[\ln(1+r/r_{\rm s})-\frac{r/r_{\rm s}}{1+r/r_{\rm s}}\right].
\end{eqnarray}

Note that the NFW fitting function Eq.~(\ref{nfwfit}) is based on $N$-body dark matter
simulations of the concordance model.
It is nontrivial to extend this assumption to modified gravity models.
However, it was shown in~\cite{Lombriser2} that the NFW profile provides equally good fits for $f(R)$ clusters as it does for the Newtonian scenario.
This was shown using $N$-body simulations of the Hu-Sawicki $f(R)$ gravity model corresponding to $\beta=\sqrt{1/6}$, which characterises only a subgroup of models of the more general chameleon model studied here.
The effects of the modifications on observables are, however, qualitatively similar between different values of the coupling strength $\beta$ and can even partially be mapped into each other, suggesting the applicability of the NFW profile.
Its validity for the full range of parameters considered in this paper may still be worthwhile checking using $N$-body simulations.
From an observational perspective, recent work by~\cite{Umetsu,Oguri} supports the consistency of the NFW profile with measurements.
Hence, even independent of the simulation results, the NFW profile could be used for the reconstruction of the lensing mass with the same motivation as introducing the gas profiles in the reconstruction of the hydrostatic mass in Appendix~\ref{sec:3dprofiles}.

We consider the virial mass of a halo within the virial radius $r_{\rm vir}$, which is related to the concentration parameter $c$ by
\begin{eqnarray}
c\equiv {r_{\rm vir}\over r_{\rm s}}.
\label{defconc}
\end{eqnarray}
The virial radius is defined such that the averaged density within this radius is
$\Delta_{\rm c}$ times the critical density.
Then the virial mass $M_{\rm vir}$ is written as
\begin{eqnarray}
M_{\rm vir}\equiv M(<r_{\rm vir})={4\pi \over 3}r_{\rm vir}^3\Delta_{\rm c}\bar\rho_{\rm c},
\label{defMvir}
 \end{eqnarray}
where $\bar\rho_{\rm c}$ is the critical density.
We use $\Delta_{\rm c}=100$ obtained
in the spherical collapse model in the cold dark matter scenario with cosmological constant $\Lambda$~\cite{NS}.
Note that the critical overdensity contrast $\Delta_{\rm c}$ generally depends on the modified gravity parameters.
For example, the authors in Ref.~\cite{LOH} found $\Delta_{\rm c}\sim 80$ in an $f(R)$ model, which is equivalent to
$\Delta_{\rm vir}\sim 300$ at redshift $z\sim 0$. Nonetheless, our final conclusion is independent
of this modification of $\Delta_{\rm c}$ because our MCMC analysis includes the parameters $M_{\rm vir}$ and $c$, which are
degenerate with $\Delta_{\rm c}$. Therefore, the change of $\Delta_{\rm c}$ only introduces shifts in the values
of $c$ and $M_{\rm vir}$.

Instead of $\rho_{\rm s}$ and $r_{\rm s}$, we can alternatively use the virial mass $M_{\rm vir}$ and concentration $c$ as the underlying fitting parameters of Eq.~(\ref{nfwfit}), from which one can determine $\rho_{\rm s}$ and $r_{\rm s}$ using the relations
\begin{eqnarray}
&&r_{\rm s}=\frac{1}{c}\left[\frac{M_{\rm vir}}{(4\pi/3)\Delta_{\rm c}\bar\rho_{\rm c}}\right]^{1/3},
\\
&&\rho_{\rm s}=\frac{M_{\rm vir}}{4\pi r_{\rm s}^3}
\left(\ln(1+c)-\frac{c}{1+c}\right)^{-1}.
\end{eqnarray}
These relations directly follow from Eqs.~(\ref{defconc}) and (\ref{defMvir}).

With the assumption of a NFW dark matter density profile of the cluster, we can derive an approximate, but analytic, solution for the radial profile of the chameleon field from Eq.~(\ref{FieldEq})~\cite{Terukina,Chameleongravity,Chameleongravity2}.
The analytic solution for Eq.~(\ref{FieldEq}) is obtained by connecting the interior solution $\phi_{\rm int}$ and the outer solution $\phi_{\rm out}$.
The interior solution is obtained when the scalar field is in the minimum of the effective potential, which corresponds to the right-hand side of Eq.~(\ref{FieldEq}).
Thus, the solution of the chameleon field can be inferred by setting $\nabla^2\phi$ to zero.
This represents the regime of the chameleon suppression of the scalar field and the chameleon field does not mediate a fifth force.
On the other hand, the outer solution is obtained when the contribution of the scalar field potential, the first term on the right-hand side of Eq.~(\ref{FieldEq}), is subdominant to the matter density and $\nabla^2\phi$.
This describes the case where the chameleon field mediates a long-range fifth force, the matter density is still large compared to the background, and the scalar field has not settled in the minimum of the effective potential.
For these two limits of the chameleon field, we find
 \begin{eqnarray}\label{ChamSol1}
 \phi(r)=\left\{
\begin{array}{ll}
\phi_{\rm s}[r/r_{\rm s}(1+r/r_{\rm s})^2]\equiv \phi_{\rm int} (\simeq 0)
& (r<r_{\rm c}) \\
-\dfrac{\beta\rho_{\rm s}r_{\rm s}^2}{M_{\rm Pl}}\dfrac{\ln(1+r/r_{\rm s})}{r/r_{\rm s}}
-\dfrac{C}{r/r_{\rm s}}+\phi_\infty \equiv\phi_{\rm out} & (r>r_{\rm c})
\end{array}
\right.,
 \end{eqnarray}
where $C$ is an integration constant and $r_{\rm c}$ is the transition scale, connecting $\phi_{\rm int}$ and $\phi_{\rm out}$.
We have furthermore defined $\phi_{\rm s}^{n+1}=(n\Lambda^{n+4}M_{\rm Pl}/\beta \rho_{\rm s})$, which represents the value of the chameleon field in the interior region, and $\phi_\infty$, the value of the scalar field at large distance from the cluster.
The chameleon field at the background is in the minimum of the effective potential, hence, we have $\phi_{\infty}^{n+1}=(n\Lambda^{n+4}M_{\rm Pl}/\beta \rho_0)$, where $\rho_0$ is the matter density at large distance from the cluster, e.g., the cosmological background density.
Due to the high density inside the cluster, $\rho_{\rm s}\gg\rho_0$, the chameleon field is strongly suppressed with $\phi_{\rm s}\ll\phi_{\infty}$.
Thus the interior solution for the scalar field Eq.~(\ref{ChamSol1}) may be approximated as $\phi_{\rm int}\simeq0$.
The integration constant $C$ and the transition scale $r_{\rm c}$
are then determined from requiring
$\phi_{\rm int}(r_{\rm c})=\phi_{\rm out}(r_{\rm c})$ and
$\phi_{\rm int}^\prime(r_{\rm c})=\phi_{\rm out}^\prime(r_{\rm c})$.
Finally we have the approximate solution
 \begin{eqnarray}\label{connection}
   &&C\simeq -\frac{\beta\rho_{\rm s} r_{\rm s}^2}{M_{\rm Pl}}\ln(1+r_{\rm c}/r_{\rm s})
 +\phi_\infty r_{\rm c}/r_{\rm s}\\
   &&\phi_\infty-\frac{\beta\rho_{\rm s} r_{\rm s}^2}{M_{\rm Pl}}(1+r_{\rm c}/r_{\rm s})^{-1}\simeq 0.
 \label{thinshell}
 \end{eqnarray}
Note that in our approximation, the chameleon field Eq.~(\ref{ChamSol1}) and the transition relations Eqs.~(\ref{thinshell}) and (\ref{connection}) do not depend on the parameters of the scalar field potential, $\Lambda$ and $n$, as we consider $\phi_{\infty}$ as the degree of freedom of the model, which, depending on the environment of a cluster, may be different from the cosmological background value of the scalar field.

From Eq.~(\ref{thinshell}), we can see that the critical radius $r_{\rm c}$, below which the chameleon field is screened, is determined by $\beta M_{\rm Pl}/\phi_{\infty}$. Hence, the smaller $\phi_\infty$ at fixed $\beta$, the larger the critical radius becomes.
As a consequence, the entire cluster can be screened.
The smaller $\beta$, the smaller the strength of the fifth force becomes.
Thus, Newtonian gravity is recovered in each of the limits $\beta = 0$ and $\phi_\infty = 0$.

In the presence of the chameleon field, the hydrostatic equilibrium Eq.~(\ref{he}) is modified by the introduction of the extra force $F_{\phi}=-(\beta/M_{\rm Pl})d\phi/dr$ on the right-hand side of the equation.
The chameleon force then modifies the mass inferred from hydrostatic equilibrium in Eq.~(\ref{HE1}) as
 \begin{eqnarray}\label{HE11}
   M_{}(<r)&=&M_{\rm thermal}(r)+M_{\rm non-thermal}(r)+M_\phi(r),
 \end{eqnarray}
where we define an extra mass
 \begin{eqnarray}
   M_{\phi}(r)&\equiv&-\frac{r^2}{G}{\beta\over M_{\rm Pl}}{d\phi(r)\over dr}
   \label{chameleonmass}
\end{eqnarray}
associated with the enhanced gravitational force due to the chameleon field.

\subsection{Inferring hydrostatic and lensing masses from observations} \label{sec:hydrostaticmass}

The thermal mass $M_{\rm thermal}$ of a cluster in Eq.~(\ref{mrrt}) is determined by its gas density, temperature, and pressure, which can be measured in X-ray and SZ observations.
In order to obtain $M_{\rm thermal}$ from observations, we reconstruct the three dimensional gas profiles using parametric fits as described in detail in Appendix~\ref{sec:3dprofiles}, which we substitute into Eq.~(\ref{mrrt}).
We assume that the gas is fully ionised and that the electron temperature is equal to $T_{\rm gas}$.
For a relaxed cluster such as Coma, used in Sec.~\ref{sec:3} to derive constraints on the chameleon model, we assume that the electrons and protons have the same temperature.
Note, however, that this assumption is nontrivial because the equipartition timescale between electrons and protons through Coulomb collisions is close to the dynamical timescale of the cluster (see, e.g.,~\cite{ClusterRev}).
Here, we use the notation $n_{\rm e}$ for the three dimensional electron number
density, which is related to the gas number density
$n_{\rm gas}$ by
\begin{eqnarray}\label{ele-gas}
  &&n_{\rm e}=\frac{2+\mu}{5}n_{\rm gas}.
  \label{nengas}
\end{eqnarray}
Similarly, we introduce the electron pressure $P_{\rm e}$, which is
related to the gas thermal pressure $P_{\rm thermal}$ by
\begin{eqnarray}\label{Pele-gas}
  &&P_{\rm e}=n_{\rm e}kT_{\rm gas}=\frac{2+\mu}{5}P_{\rm thermal}.
\label{xp}
\end{eqnarray}

With the definitions in Eqs.~(\ref{nengas}) and (\ref{xp}) and the reconstructed 3-dimensional temperature, electron density, and pressure profiles from Appendix~\ref{sec:3dprofiles}, we can now determine the thermal mass profile of the cluster.
From X-ray observations, we infer
\begin{eqnarray}
M_{\rm thermal}=-\frac{kT_{\rm gas}^{\rm (X)}r}{\mu m_{\rm p}G}\left(
\frac{d\ln n_{\rm e}^{\rm (X)}}{d\ln r}+\frac{d\ln T_{\rm gas}^{\rm (X)}}{d\ln r}\right)
\label{thermalX}
\end{eqnarray}
and similarly, from the SZ observations, we obtain
\begin{eqnarray}
M_{\rm thermal}=-\frac{r^2}{G\rho_{\rm gas}^{\rm (X)}}\frac{dP_{\rm thermal}^{\rm (SZ)}}{dr}.
\label{thermalSZ}
\end{eqnarray}
With this reconstruction, we can directly compare the two mass profiles with the lensing mass
\begin{eqnarray}\label{NFWmass}
M_{\rm WL}=4\pi \rho_{\rm s} r_{\rm s}^3\biggl[
  \ln(1+r/r_{\rm s})-{r/r_{\rm s}\over 1+r/r_{\rm s}}
\biggr],
\label{lensingM}
\end{eqnarray}
which is obtained by integration over the NFW density profile in Eq.~(\ref{nfwfit}), assuming that $\phi/M_{\rm Pl}\ll1$ such that the lensing potential is related to the matter distribution by the standard Poisson equation.

In the presence of a non-thermal pressure, Eqs.~(\ref{thermalX}) and (\ref{thermalSZ}) are modified according to Eq.~(\ref{grg}) with the mass inferred from X-ray by
\begin{eqnarray}
&&M_{\rm thermal}^{}+ M_{\rm non-thermal}^{}
\nonumber
\\
&&~~~~~~~~~~
=-\frac{kT_{\rm gas}^{\rm (X)}r}{\mu m_{\rm p}G}\left(
\frac{d\ln n_{\rm e}^{\rm (X)}}{d\ln r}+\frac{d\ln T_{\rm gas}^{\rm (X)}}{d\ln r}\right)
-{r^2\over G \rho_{\rm gas}^{\rm (X)}}{d\over dr}\left(
{g\over 1-g}n_{\rm gas}^{\rm (X)}kT_{gas}^{\rm (X)}\right),
\label{nonthmX}
\end{eqnarray}
whereas a combination of SZ and X-ray infers
 \begin{eqnarray}
M_{\rm thermal}^{}+ M_{\rm non-thermal}^{}=
-\frac{r^2}{G\rho_{\rm gas}^{\rm (X)}}\frac{dP_{\rm thermal}^{\rm (SZ)}}{dr}
-{r^2\over G \rho_{\rm gas}^{\rm (X)}}{d\over dr}\left(
{g\over 1-g}P_{\rm thermal}^{\rm (SZ)}\right).
\label{nonthmSZ}
\end{eqnarray}

To derive our constraints in Sec.~\ref{sec:3}, we will assume hydrostatic equilibrium, Eq.~(\ref{he}), and thus require
\begin{equation}
 M_{\rm thermal} + M_{\rm non-thermal} + M_{\phi} \equiv M_{\rm WL}, \label{hydroconstraints}
\end{equation}
where $M_\phi$ is the chameleon contribution described in Eq.~(\ref{chameleonmass}) and $M_{\rm non-thermal} \ll M_{\rm thermal} + M_{\phi}$.
We refer the reader to Sec.~\ref{sec:systematics1} for an analysis on the validity of the hydrostatic equilibrium assumption in the case of the Coma cluster.

%
\section{Application to the Coma cluster} \label{sec:3}

Having established the notion of hydrostatic and lensing mass, and having described the effects from
the presence of a chameleon field on the relation between the two in Sec.~\ref{sec:2}, we now analyse
constraints on the chameleon gravity model by confronting our predictions with observations of the
Coma cluster.
We chose to work with the Coma cluster as it is a relaxed system, where the non-thermal pressure is expected to be subdominant 
(e.g. \cite{Veritas} and also see Sec.~\ref{sec:systematics1}) and which has been well measured through a range of different observations ~\cite{Veritas,Sreekumar,Reimer,Ackermann}.
The contribution of non-thermal pressure can also be assumed small in modified gravity~\cite{Arnold}.
Ref.~\cite{Pimbblet} has recently pointed out that the cluster may not be very typical: its X-ray temperature and star formation rate is high but the kinematic features like substructure and velocity dispersion are not conspicuous. The authors urge caution in using Coma cluster as a $z \sim 0$ baseline cluster in galaxy evolution studies.
On the other hand, according to references \cite{Zhang, planckearly}, the Coma cluster is in agreement with scaling relations obtained from typical cluster samples. We cannot exclude that extraordinary features of the cluster may affect our conclusions. However, our constraints rely only on the observed distribution of gas and dark matter and we allow a number of degrees of freedom to phenomenologically model these distributions, finding good agreement of our fits with the observational data. We also carefully examine a dynamical equilibrium model of the Coma cluster.
Note that our method applies to any cluster which is in hydrostatic equilibrium, and is not restricted to the Coma cluster.
This section is organized as follows:
In Sec.~\ref{sec:mcmc}, we first assume hydrostatic equilibrium and model the effect from chameleon
gravity using the analytic scalar field solution described in Sec.~\ref{sec:chameleon} to derive and
compare the theoretical gas distribution profiles with the corresponding observations of the Coma cluster.
Then we simultaneously fit for the observed X-ray surface brightness, the X-ray temperature,
the SZ effect, and the WL profile based on a parametric fit for the electron number density and
the NFW profile.
We obtain competitive constraints on the chameleon model.
In particular, our method provides the currently strongest cosmological constraint on $f(R)$
gravity (see Fig.~\ref{fig:mg_limit}).
In Sec.~\ref{sec:systematics}, we then analyse the validity of the hydrostatic equilibrium
assumption of gas in the Coma cluster and study the potential systematic effects on the
reconstructed mass profiles as well as possible errors induced by non-spherically symmetric
features of the cluster.
%


\subsection{Constraints on the model parameters from an MCMC analysis} \label{sec:mcmc}
We estimate the 3-dimensional profiles of the temperature, electron density, and pressure from observations of the X-ray temperature, surface brightness, and SZ effect employing the parametric fits described in Appendix~\ref{sec:3dprofiles}, as well as the lensing mass, for which we use a NFW profile.
In hydrostatic equilibrium, the hydrostatic mass can then be inferred from any combination of two of these profiles.
Here, we choose to work with the electron number density Eq.~(\ref{nf}) and the NFW profile Eq.~(\ref{NFWmass0}), and perform an MCMC analysis of the model parameter space, including the chameleon model parameters $\beta$ and $\phi_{\infty}$ discussed in Sec.~\ref{sec:chameleon}.

\subsubsection{Method}

We write the hydrostatic equilibrium equation as
 \begin{eqnarray}
   \frac{1}{\rho_{\rm gas}(r)}\frac{P_{\rm thermal}(r)}{dr}
   =-\frac{GM(<r)}{r^2}-\frac{\beta}{M_{\rm Pl}}\frac{d\phi(r)}{dr}
 \label{herp}
 \end{eqnarray}
and assume that the equation of state for the gas is given by $P_{\rm thermal} = n_{\rm gas}kT_{\rm gas}$, which is equivalent to $P_{\rm e}=n_{\rm e}kT_{\rm gas}$, where the electron temperature equals to $T_{\rm gas}$.
Integration of Eq.~(\ref{herp}) yields
 \begin{eqnarray}
   P_{\rm thermal}(r)&=&P_{\rm thermal,0}+\mu m_{\rm p}\int^r_0 n_{\rm e}(r)
\left(-\frac{GM(<r)}{r^2}-\frac{\beta}{M_{\rm Pl}}\frac{d\phi(r)}{dr} \right)dr. \label{pthermal}
 \end{eqnarray}
Hereby, $P_{\rm thermal,0}$ is an integration constant equal to the thermal gas pressure at $r=0$, which can be written as $P_{\rm thermal,0}=n_{\rm gas,0}kT_0$, where $n_{\rm gas,0}$ and $T_0$ are the thermal gas number density and the gas (electron) temperature at $r=0$, respectively.
We adopt Eq.~(\ref{nf}) to describe the electron number density $n_{\rm e}(r)$ and the NFW profile Eq.~(\ref{nfwfit}) for the matter density which determines the cluster mass profile $M(<r)$ in Eq.~(\ref{NFWmass0}).
Note from Eq.~(\ref{ele-gas}) that $n_{\rm gas,0}$ is expressed by $n_0$ as $n_{\rm gas,0}=5n_0/(2+\mu)$.

The NFW density profile is specified by the virial mass $M_{\rm vir}$ and the concentration parameter $c$.
The configuration of the scalar field is given by specifying the parameters $\beta$ and $\phi_\infty$.
Including the parameters for the electron number density, the
complete list of parameters we analyse in our MCMC study becomes $T_0, n_0, b_1, r_1$, $M_{\rm vir}, c, \beta, \phi_\infty$, where $r_1$ and $b_1$ determine a characteristic scale and slope, respectively, for $n_{\rm e}(r)$ in Eq.~(\ref{nf}).
Once these parameters are specified, we can compute the projected gas profiles in Eqs.~(\ref{xt}), (\ref{sb}), and (\ref{sz}), which are then compared with the observational data from the X-ray surface brightness and temperature, and the SZ observations.

We estimate the ``goodness-of-fit" by computing the chi-squared distribution
 \begin{eqnarray}\label{chitotal}
   \chi^2(M_{\rm vir},c,T_0,n_0,b_1,r_1,
\beta,\phi_\infty)=\chi_{\rm XT}^2+\chi_{\rm SB}^2+\chi_{\rm SZ}^2+\chi_{\rm WL}^2,
 \end{eqnarray}
where
 \begin{eqnarray}
   \chi_{\rm XT}^2&=&\sum_{i}^{}\frac{(T_{\rm X}(r_{\perp,i})-T_{{\rm X},i}^{\rm obs.})^2}
                 {(\Delta T_{{\rm X},i}^{\rm obs.})^2},
\\
   \chi_{\rm SB}^2&=&\sum_{i}^{}\frac{(S_{\rm X}(r_{\perp,i})-S_{{\rm X},i}^{\rm obs.})^2}
                 {(\Delta S_{{\rm X},i}^{\rm obs.})^2},
\\
   \chi_{\rm SZ}^2&=&\sum_{i}^{}\frac{(y_{}(r_{\perp,i})-y_{i}^{\rm obs.})^2}
                 {(\Delta y_{i}^{\rm obs.})^2},
\\
   \chi_{\rm WL}^2&=&\frac{(M_{\rm vir}-M_{\rm WL})^2}{(\Delta M_{\rm WL})^2}
                 +\frac{(c_{}-c_{\rm WL})^2}{(\Delta c_{\rm WL})^2}.
 \end{eqnarray}
Here, $T_{\rm X}(r_{\perp,i})$ and $T_{{\rm X},i}^{\rm obs.}$ are the theoretical and observed X-ray temperatures, and $\Delta T_{{\rm X},i}^{\rm obs}$ refers to the observational error.
We adopt the analogous notation for the surface brightness $S_{\rm X}$ and the $y$-parameter, defined by the SZ temperature as $\Delta T_{\rm SZ}/T_{\rm CMB}\equiv -2y$.
In addition, $M_{\rm WL}$ and $c_{\rm WL}$ are the observed virial mass and the concentration parameter from weak lensing, respectively.

For the X-ray temperature profile, we use the {\it XMM-Newton} data reported
in \cite{Coma_3} for the inner region and {\it Suzaku} data reported in \cite{Coma_4}
for the outer region.
For the X-ray surface brightness profile, we use the {\it XMM-Newton} data reported in \cite{Coma_2}
and for the SZ pressure profile, we use the \emph{Planck} measurements~\cite{Coma_5}.
Finally, we use the WL measurement of the Coma cluster reported by Okabe \emph{et al.} \cite{Coma_7},
who adopt a NFW fit in their analysis to obtain a virial mass of the Coma cluster of $M_{\rm vir}=8.92^{+20.05}_{-5.17}$$\times10^{14}h^{-1} M_{\odot}$ and a concentration of $c=3.5^{+2.57}_{-1.79}$ with virial overdensity $\Delta_{\rm c}= 100$.

In our likelihood analysis, we assume that the information contained in each data point is independent of the other data points, i.e., that there is no correlation between these four observations.
This could be an over-simplification.
These four observations are based on different measurement principles, and the X-ray, SZ effect, and WL observations are obtained at different wavelengths.
On the other hand, the information contained in the data comes from the same astrophysical object, and thus the systematic errors might be correlated.
For instance, the clumpiness of the cluster and other non-spherically symmetric features would introduce a correlated systematic error between the data sets.
We do not take into account such correlations in our analysis and leave it for future work to address these observational issues in more detail.
See, however, Sec.~\ref{sec:systematics} for a discussion of these effects. 
We also note that the covariance of errors is not taken into account in our analysis because it is not available to us. 
For now, we assign a 5\% systematic error to the measurement error of the X-ray surface brightness.

\subsubsection{MCMC analysis} \label{mcmcanalysis}

We perform an MCMC analysis with the 8 model parameters $T_0, n_0, b_1, r_1$, $M_{\rm vir}, c, \beta_2$, and $\phi_{\infty,2}$, which completely describe the X-ray temperature and surface brightness profiles, the SZ effect, and the WL mass profile as well as the chameleon modified gravity model. We re-normalise the parameters $\beta_2=\beta/(1+\beta)$ and $\phi_{\infty,2}=1-\exp(-\phi_\infty/10^{-4}M_{\rm Pl})$ (instead of $\beta$ and $\phi_\infty$) as $\beta_2$ and $\phi_{\infty,2}$ span the complete available parameter space of $\beta$ and $\phi_\infty$ in the interval $[0,1]$.
Note, however, that some of the approximations made in Sec.~\ref{sec:2} do not hold in the extreme limits of $\phi_{\infty,2}\rightarrow1$ and $\beta_2\rightarrow1$.
For our analysis, we use the MCMC module included in the {\sc cosmomc}~\cite{cosmomc} package, which employs a Metropolis-Hastings~\cite{metropolis,hastings} sampling algorithm.
We require a Gelman-Rubin statistic~\cite{gelman} of $\mathcal{R} - 1 < 0.03$ to ensure convergence of our runs.

\begin{figure}
 \begin{tabular}{cc}
  \begin{minipage}[t]{.5\hsize}
   \includegraphics[width=\hsize,clip]{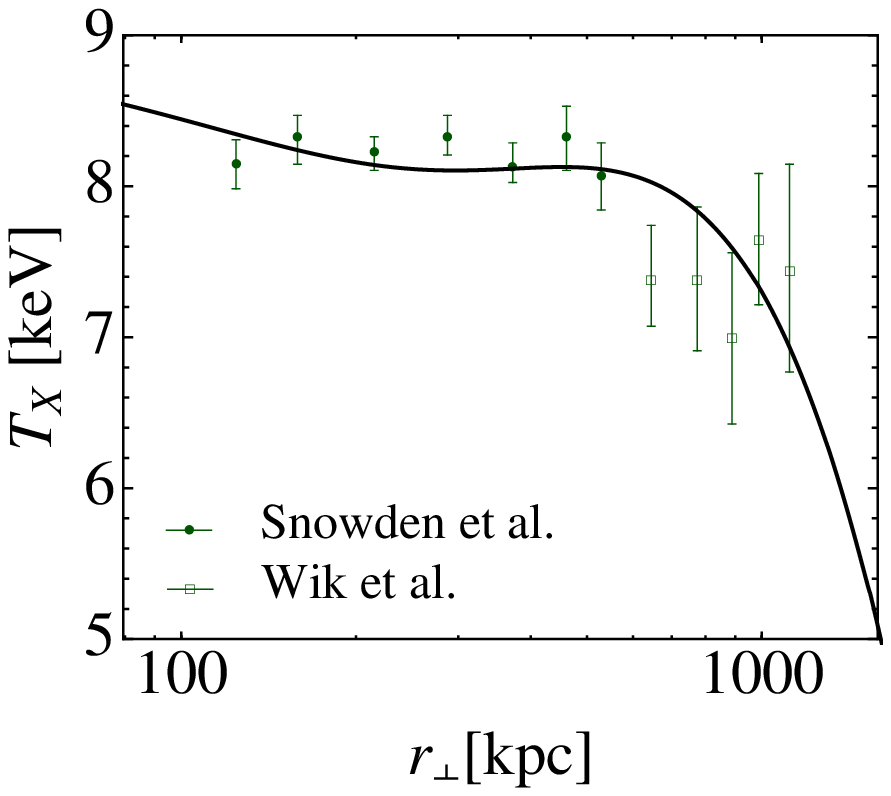}
  \end{minipage}
  \begin{minipage}[t]{.5\hsize}
   \includegraphics[width=\hsize,clip]{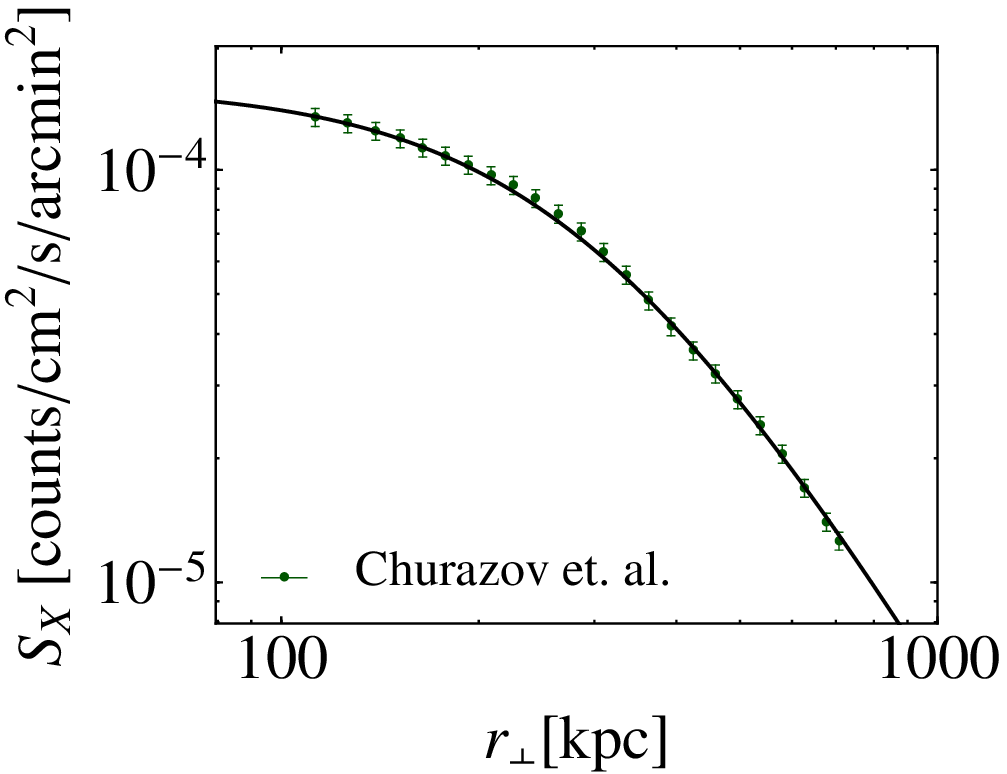}
  \end{minipage}
 \end{tabular}
 \includegraphics[width=0.5\hsize,clip]{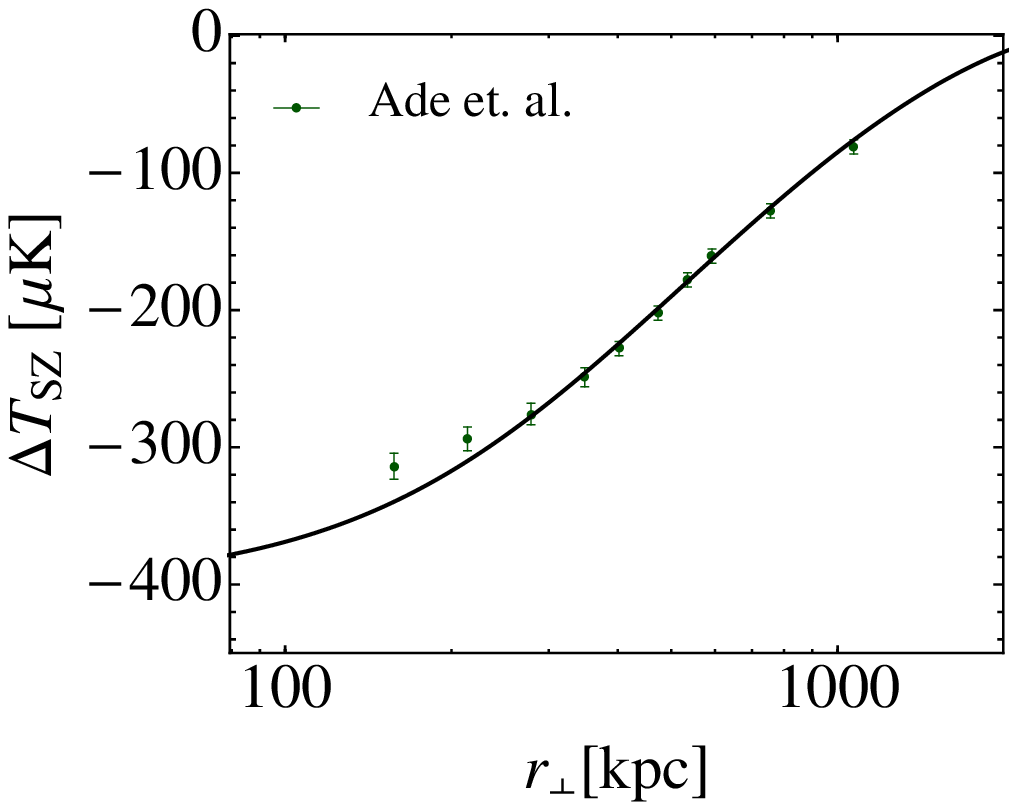}
\caption{\label{fig:mcmc1bestfit}
X-ray temperature (top-left), surface brightness (top-right), and SZ effect (bottom).
The best-fit values of the chameleon model parameters are 
($\beta$, $\phi_{\infty}$)=($15$, $4\times 10^{-4}M_{\rm Pl}$), 
where the model parameters characterising the profiles are given in Table~\ref{table1}.
In the data analysis, we use the data points included within the radial range $100~ {\rm kpc}< r_\perp < 1~{\rm Mpc}$
and fit them using the model parameters $T_0, n_0, b_1, r_1$, $M_{\rm vir}, c$ in the Newtonian case and in addition $\beta_2$ and $\phi_{\infty,2}$ in the chameleon scenario.
Note that the best-fits of the Newtonian and chameleon cases almost overlap.
\label{fig:Coma_T_MCMC}
}
\end{figure}

\begin{table}[t]
\caption{
Best-fit values and 1-dimensional marginalised constraints ($95$\% CL)
for the model parameters ($T_0, n_0, b_1, r_1$, $M_{\rm vir}, c$)
characterising the gas and dark matter profiles obtained from an MCMC analysis of the joint observational data sets.
\label{table1}
}
\begin{center}
\renewcommand{\arraystretch}{1.5}
\begin{tabular}{c|c|c}
\hline\hline
~~~~parameter~~~~ & ~~~~Newtonian gravity~~~~ & ~~~~Modified gravity~~~~\\
\hline
$M_{\rm vir}$ & ${2.57}^{+0.97}_{-0.54} ~10^{15}~M_\odot$  & ${2.46}^{+1.33}_{-0.61} ~ 10^{15}~M_\odot$\\
$c$                & ${2.56}^{+0.49}_{-0.52}$               & ${2.64}^{+0.72}_{-0.7}$\\
$n_0$ & ${2.33}^{+0.22}_{-0.17} ~10^{-3}{\rm /cm^3}$ & ${2.34}^{+0.21}_{-0.19} ~ 10^{-3}{\rm /cm^{3}}$\\
$b_1$              & ${-0.921}^{+0.089}_{-0.109}$            & ${-0.915}^{+0.085}_{-0.107}$\\
$r_1$    & ${3.02}^{+0.54}_{-0.47}~ 10^2~{\rm kpc}$     & ${2.99}^{+0.56}_{-0.45}~ 10^2~{\rm kpc}$\\
$T_0$    & ${11.2}^{+0.76}_{-0.84} ~{\rm keV}$               & ${11.3}^{+0.79}_{-0.9}~{\rm keV}$\\
\hline\hline
\end{tabular}
\renewcommand{\arraystretch}{1.}
\end{center}
\end{table}

In Fig.~\ref{fig:mcmc1bestfit}, we compare the overall best-fit curves for the chameleon gravity model (dashed) and Newtonian gravity (solid) from the combination of all of the observational data sets, i.e., minimising $\chi^2$ in Eq.~(\ref{chitotal}).
The corresponding best-fit parameter values are listed in Table~\ref{table1} along with the 1-dimensional marginalised 95\% confidence levels (CL).
We show the 2-dimensional marginalised contours of the different combinations between the model parameters for the Newtonian case, i.e., where we have fixed $\beta=0$ and $\phi_\infty=0$, in Fig.~\ref{fig:mcmc1x}.
The best fit in this case yields a reduced $\chi^2$ of $\chi^2/{\rm d.o.f.}=32/41$. 
In Fig.~\ref{fig:mcmcx}, we show the analogous constraints for the model parameters of the chameleon modified gravity scenario.
The best fit in this case yields a good reduced $\chi^2$ of $\chi^2/{\rm d.o.f.}=32/39$. 
We refer to Sec.~\ref{sec:systematics} for a discussion of possible sources of systematic error that have not been taken into account in this analysis.

%

\begin{figure}[t]
\begin{center}
  \includegraphics[width=0.6\hsize,clip]{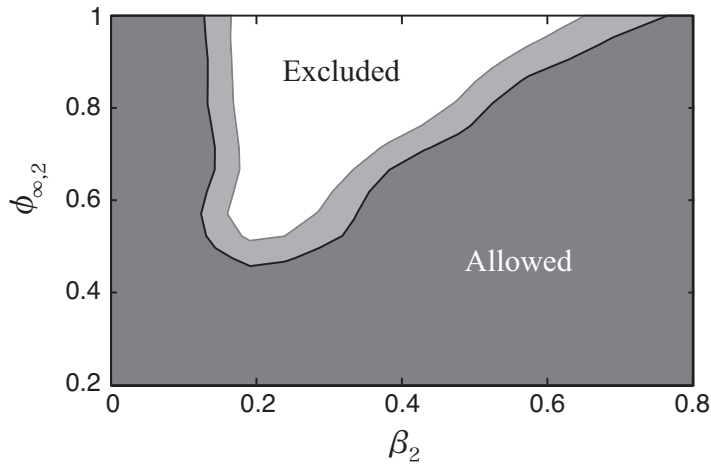}
\caption{
95\% (deep gray region) and 99\% CL (pale gray region) contours for the chameleon model parameters $\beta_2=\beta/(1+\beta)$ and $\phi_{\infty,2}=1-\exp(-\phi_\infty/10^{-4}{\rm M_{Pl}})$, obtained from the MCMC analysis of the 8 model parameters, $T_0, n_0, b_1, r_1,  M_{\rm vir}, c, \beta_2$, and $\phi_{\infty,2}$, using the joint set of X-ray, SZ, and WL data.
The shaded region is the allowed region.
\label{fig:Coma_phinf-beta2x}}
\end{center}
\end{figure}

Finally, in Fig.~\ref{fig:Coma_phinf-beta2x}, we show the 2-dimensional marginalised contours of the parameters $\beta_2$ and $\phi_{\infty,2}$.
Note that the lower shaded region is the allowed region.
We recall that $\beta$ describes the strength of the chameleon fifth force and $\phi_\infty$ determines the efficiency of the chameleon screening, and we introduced the parameters $\beta_2=\beta/(1+\beta)$, which we mapped into $\phi_{\infty,2}=1-\exp(-\phi_\infty/10^{-4}M_{\rm Pl})$ instead of $\beta$ and $\phi_\infty$ to describe the entire parameter space of the chameleon modification.
Newtonian gravity is recovered in both limits of $\beta_2=0$ and $\phi_{\infty, 2}=0$.

The boundaries in Fig.~\ref{fig:Coma_phinf-beta2x} can be understood by considering the phenomenology of the chameleon modification.
At large $\beta$, if the chameleon field is not screened, the extra chameleon force
reduces the hydrostatic mass compared to the Newtonian mass estimate and it becomes inconsistent with the lensing mass (see Sec.~\ref{sec:systematics}).
This causes a tension in the desired parameter values when fitting the joint set of observations and places constraints on the chameleon modification.
On the other hand, the chameleon force contributes only outside of the critical radius $r_{\rm c}$, which is determined by Eq.~(\ref{thinshell}) as
\begin{equation}
1 + \frac{r_{\rm c}}{r_{\rm s}} = \frac{\beta \rho_{\rm s} r_{\rm s}^2}{M_{\rm Pl} \phi_\infty}.
\label{cp}
\end{equation}
Due to the chameleon suppression mechanism, Newtonian gravity is recovered below $r_{\rm c}$.
To put a useful constraint on the chameleon force, $r_{\rm c}$ must be smaller than the size of the cluster, which is about 1~Mpc.
More precisely, with increasing $\beta M_{\rm Pl} /\phi_\infty$, the transition scale $r_{\rm c}$ becomes large and eventually surpasses the size of the cluster, in which case the chameleon mechanism completely screens the fifth force within the cluster.
At this point, no further constraints on the chameleon model can be obtained.
This implies that there is an upper bound on $\beta M_{\rm Pl} /\phi_\infty$, which can be constrained.
In the opposite limit, when $\beta$ is small, the fifth force is weak and the modifications become consistent with the observations within the given errors.
Hence, at low $\beta_2$ in Fig.~\ref{fig:Coma_phinf-beta2x} the chameleon scalar field amplitude $\phi_{\infty,2}$ is unconstrained.

With the minimal scalar field in the background, $-\Lambda^{n+4} \simeq n^{-1} \beta\,\bar{R}_0\,\phi_{\infty}^{n+1} M_{\rm Pl}$,
the Compton wavelength of the background scalar field today, assumed to be $\phi_{\infty}$ here, becomes~\cite{Lombriser5}
\begin{equation}
 m_{\infty}^{-1} \simeq \left[ \frac{\beta \, \bar{R}_0}{n+1} \frac{M_{\rm Pl}}{\phi_{\infty}} \right]^{-1/2} \sim \left[ 10^{-6} \frac{\beta}{n+1} \frac{M_{\rm Pl}}{\phi_{\infty}} \right]^{-1/2}~{\rm Mpc}.
\end{equation}
Whereas the chameleon mechanism suppresses the scalar field on scales below $r_{\rm c}$, on scales larger than the Compton wavelength $m_{\infty}^{-1}$, modifications of gravity are Yukawa suppressed.
With Solar System tests requiring that $\phi_{\infty} \lesssim 10^{-6}\beta$~\cite{HS,Lombriser5} and with $n\sim\mathcal{O}(1)$, one obtains $m_{\infty}^{-1}\sim{\rm Mpc}$.
Hence, requiring Solar System tests to be satisfied, standard gravity is recovered on scales beyond $\mathcal{O}(1)~{\rm Mpc}$ (cf.~\cite{WHK}).
Since we only use observations on scales smaller than $1~{\rm Mpc}$ and constraints are weaker than the local bounds, we can safely ignore the Yukawa suppression.

\subsubsection{Constraint on $f(R)$ gravity} \label{sec:fRconstraints}

Our constraints have important implications for $f(R)$ gravity~\cite{Starobinski,HS,Tsujikawa}, which corresponds to a subset of our models with a particular choice of the coupling constant $\beta=\sqrt{1/6}$.
In $f(R)$ gravity, the Einstein-Hilbert action is supplemented by a free nonlinear function $f(R)$ of the Ricci scalar R,
\begin{eqnarray}
S={1\over 16\pi G}\int d^4x\sqrt{-g}(R+f(R))+\int d^4x\sqrt{-g} L_{\rm m},
\end{eqnarray}
where $L_{\rm m}$ is the matter Lagrangian.
Here, we adopt the particular expression $f(R)=-m^2c_1/c_2+(m^2c_1/c_2^2)(R/m^2)^{-\tilde n}$ of the Hu-Sawicki model~\cite{HS}, where $\tilde n$, $m$, $c_1$ and $c_2$ are constant model parameters.
Note that $m^2c_1/c_2/2$ can be chosen such that the modification exhibits an effective cosmological constant and mimics the expansion history of the concordance model.
Hence, we specify $m^2=\Omega_{\rm m}H_0^2$ and $c_1/c_2=6\Omega_\Lambda/\Omega_{\rm m}$, where $\Omega_{\rm m}$ and $H_0$ are the matter density parameter and the Hubble parameter at the present epoch, respectively, and $\Omega_\Lambda\equiv1-\Omega_{\rm m}$.
Furthermore, we have $\tilde n c_1/c_2^2=-f_{R0}[3(1+4\Omega_\Lambda/\Omega_{\rm m})]^{\tilde n+1}$, where we introduced the model parameter $f_{R0}$, which is the value of $d f(R)/ dR$ at present time and at the background.
The $f(R)$ modification can be related to the chameleon field $\phi$ via
\begin{eqnarray}\label{frToCham}
  &&f_{R} = -\sqrt{\frac{2}{3}}\frac{\phi}{M_{\rm Pl}}
\end{eqnarray}
and hence, assuming that the Coma cluster is isolated such that $\phi_{\infty}$ corresponds to the background scalar field value, we have
$f_{R0} = -\sqrt{2/3}(\phi_\infty/M_{\rm Pl})$.
{}From the 2-dimensional contours of $(\beta_2,\phi_{\infty,2})$ in Fig.~\ref{fig:Coma_phinf-beta2x}, we therefore estimate an upper bound on $f(R)$ gravity of $\phi_\infty\simlt7\times 10^{-5}M_{\rm Pl}$
or equivalently, $|f_{\rm R0}|\simlt6\times 10^{-5}$ at 95\% CL.

We emphasise that this is a competitive result with the bounds on $f(R)$ gravity obtained from cosmology such as from the abundance of clusters~\cite{Ob1,Lombriser3,Lombriser4} (see Fig.~\ref{fig:mg_limit}) and the current constraints from redshift-space distortions in the large scale structure of galaxies \cite{YNHNS}.
Note that in the case of $\tilde n=1$, the value of $|f_{R0}|$ is related to the Compton wavenumber of the scalar field $k_C$ by
\begin{eqnarray}\label{frToCham}
  &&k_C\simeq 0.04\left(10^{-4}\over |f_{R0}|\right)^{1/2}~h{\rm Mpc}^{-1}.
\end{eqnarray}
Then, $ |f_{\rm R0}|\simlt 6\times 10^{-5}$ can be rephrased as $k_C\simlt0.05~h{\rm Mpc}^{-1}$.

Note that the assumption that the Coma cluster is an isolated system is nontrivial.
It is well known that on large scales, the cluster is connected to a network of filaments~\cite{gregory,west}.
Hence, $\phi_\infty$ or $f_{R0}$ should really be understood as the scalar field value in the mean density environment within a large radius around the Coma cluster, which we expect to be close to the background value~\cite{Rines}.
This interpretation does not differ from approaches taken to derive the constraints reported in Fig.~\ref{fig:mg_limit}.
Another possible effect which may be introduced by the environment could be a large-scale non-spherically symmetric feature as discussed in Sec.~\ref{sec:systematics2}.

\subsection{Systematic effects} \label{sec:systematics}
%

So far we have assumed hydrostatic equilibrium of the gas and a spherically symmetric matter distribution.
We therefore devote the remainder of this section to discuss the systematic errors that can be introduced in our analysis due to deviations of the hydrostatic equilibrium (Sec.~\ref{sec:systematics1}) and to adumbrate the systematics caused by the presence of non-spherically symmetric features (Sec.~\ref{sec:systematics2}).

\subsubsection{Invalidity of hydrostatic equilibrium} \label{sec:systematics1}

By employing the assumption of hydrostatic equilibrium in our analysis of the model parameter space, we have supposed that for the Coma cluster, the hydrostatic masses inferred from temperature and density, and that from pressure and density, are consistent, as well as that the two hydrostatic masses are also consistent with the lensing mass.
Here, we test the validity of hydrostatic equilibrium within Newtonian gravity by comparing the different mass estimates, and study the effects of introducing non-thermal pressure.

\begin{figure}
 \begin{tabular}{cc}
  \begin{minipage}[t]{.5\hsize}
   \includegraphics[width=\hsize,clip]{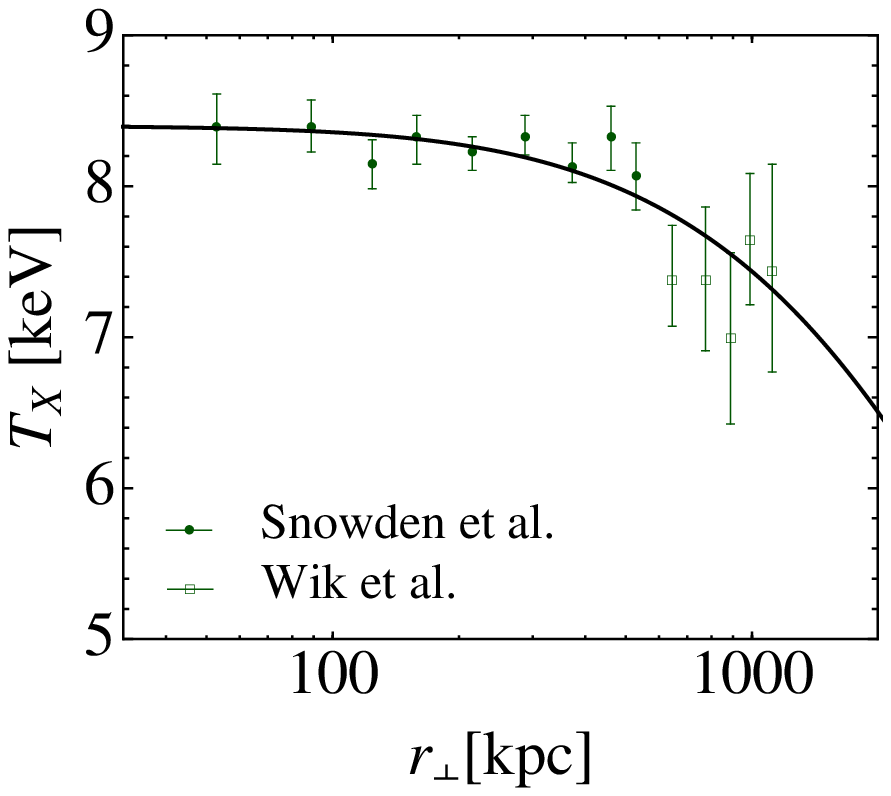}
  \end{minipage}
  \begin{minipage}[t]{.5\hsize}
   \includegraphics[width=\hsize,clip]{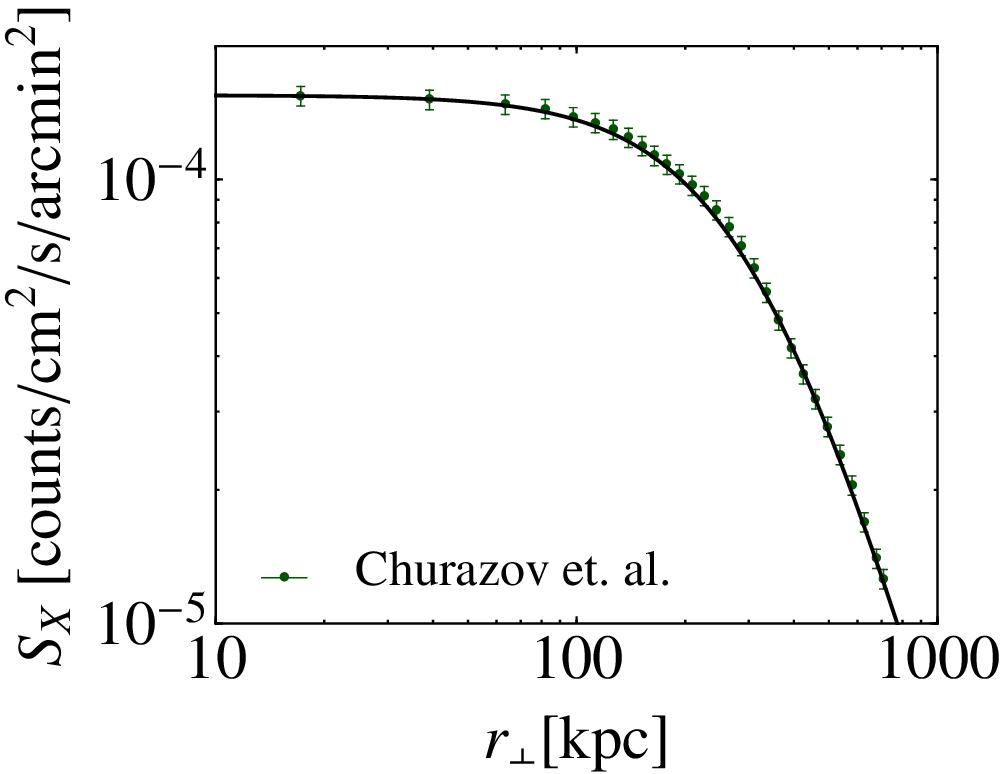}
  \end{minipage}
 \end{tabular}
   \includegraphics[width=0.5\hsize,clip]{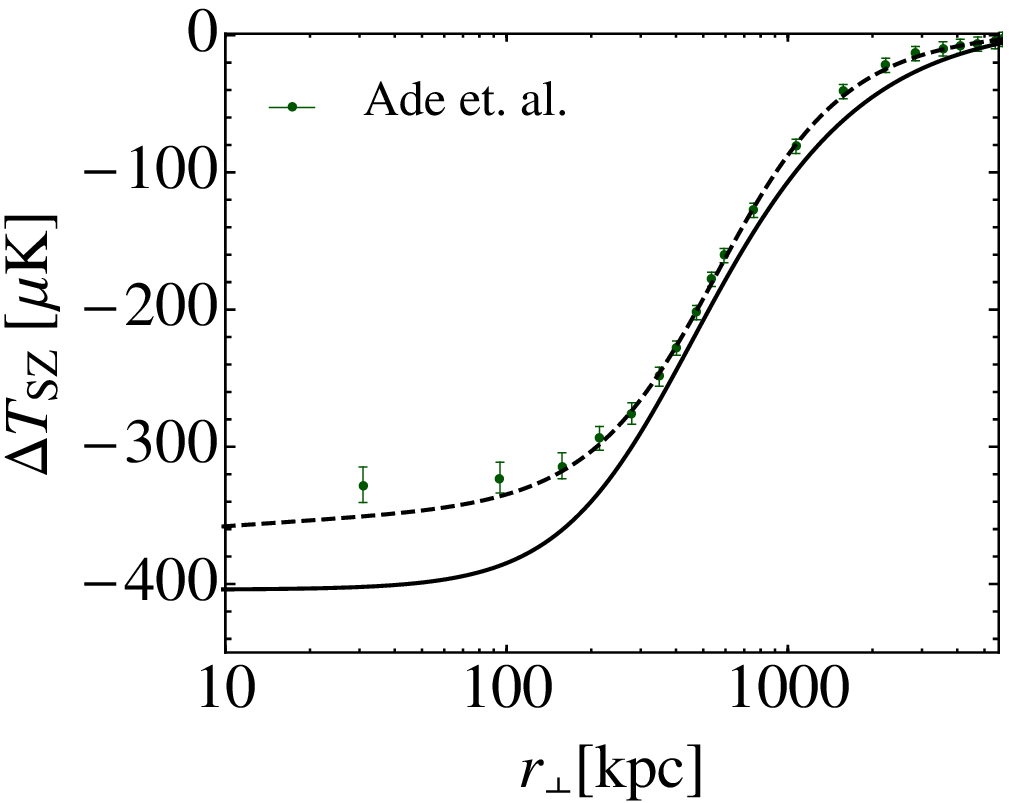}
\caption{
\emph{Top left panel}:
Radial gas temperature profile of the Coma cluster.
The circles and boxes represent the the data points and errors from the {\it XMM-Newton} measurements by Snowden \emph{et al.}~\cite{Coma_3} and the {\it Suzaku} measurements by Wik \emph{et al.}~\cite{Coma_4}, respectively.
The solid curve is the projected emission weighted temperature Eq.~(\ref{xt}), using the fitting functions Eqs.~(\ref{tf}) and (\ref{nf}) for the 3-dimensional temperature and electron density profiles with best-fit parameter values 
$(T_0, A, r_0, b_0)=(8.6~{\rm keV}, 0.082, 3.9~{\rm Mpc}, -5.3)$ and $(n_0, r_1, b_1)=(2.3\times 10^{-3}{\rm cm}^{-3}, 0.34~{\rm Mpc}, -1)$, 
respectively, to the joint X-ray data.
\emph{Top right panel}:
Radial surface brightness profile of the Coma cluster.
The data points represent the {\it XMM-Newton} measurements by Churazov \emph{et al.}~\cite{Coma_2}.
The error bars in the original data, which only account for the Poisson noise contribution, are small.
We assign a systematic error of 5\% to each data point to take into account clumpiness and other non-spherically symmetrical features of the cluster.
The solid curve is the surface brightness profile Eq.~(\ref{sb}), using the fitting functions Eqs.~(\ref{nf}) and (\ref{tf}) for the 3-dimensional electron density profile temperature profile with best-fit parameter values 
$(T_0, A, r_0, b_0)=(8.6~{\rm keV}, 0.082, 3.9~{\rm Mpc}, -5.3)$ and $(n_0, r_1, b_1)=(2.3\times 10^{-3}{\rm cm}^{-3}, 0.34~{\rm Mpc}, -1)$, 
respectively, to the joint X-ray data.
\emph{Bottom panel}:
Radial Sunyaev-Zel'dovich CMB temperature profile.
The data points represent the {\it Planck} measurements by Ade \emph{et al.}~\cite{Coma_5}.
The dashed curve is the SZ effect Eq.~(\ref{sz}), using the fitting function Eq.~(\ref{pf}) for the 3-dimensional pressure profile with best-fit parameter values
$(P_0, b_3, b_4, b_5, r_4)=(1.1\times 10^{-2}~{\rm keV/cm^{3}}, 0.14, 2.2, 1.1, 0.53~{\rm Mpc})$.
The solid curve is the best fit model from the joint X-ray observations.
\label{fig:Coma1}
}
\end{figure}

In the top left and top right panel of Fig.~\ref{fig:Coma1},
we compare the observed X-ray temperature and surface brightness, respectively, against the corresponding best fit curves, which are obtained by fitting the projected profiles of Eqs.~(\ref{xt}) and (\ref{sb}) with Eqs.~(\ref{tf}) and (\ref{nf}) to the combined X-ray data.
Note that in the top right panel, for each data point, we have assigned a 5\% systematic error on top of the measured errors.
The measured errors for the X-ray surface brightness are extremely small because they only include the Poisson noise contribution.
Systematic errors can be introduced from the clumpiness and the non-spherical symmetry of the gas distribution and should be taken into account (see Sec.~\ref{sec:systematics2}).

The bottom panel of Fig.~\ref{fig:Coma1} shows the SZ observations by the {\it Planck} satellite reported in~\cite{Coma_5}, which
we compare
with the two different best-fit curves.
The dashed curve
is the best fit obtained by fitting the SZ profile Eq.~(\ref{sz}) with Eq.~(\ref{pf}) and the solid curve is the best fit to the joint X-ray temperature and surface brightness data, i.e., with the same parameter values used in the top left and top right panels of Fig.~\ref{fig:Coma1}.
Note the deviation between the two curves.
%

\begin{figure}[t]
\begin{center}
  \includegraphics[width=0.60\hsize,clip]{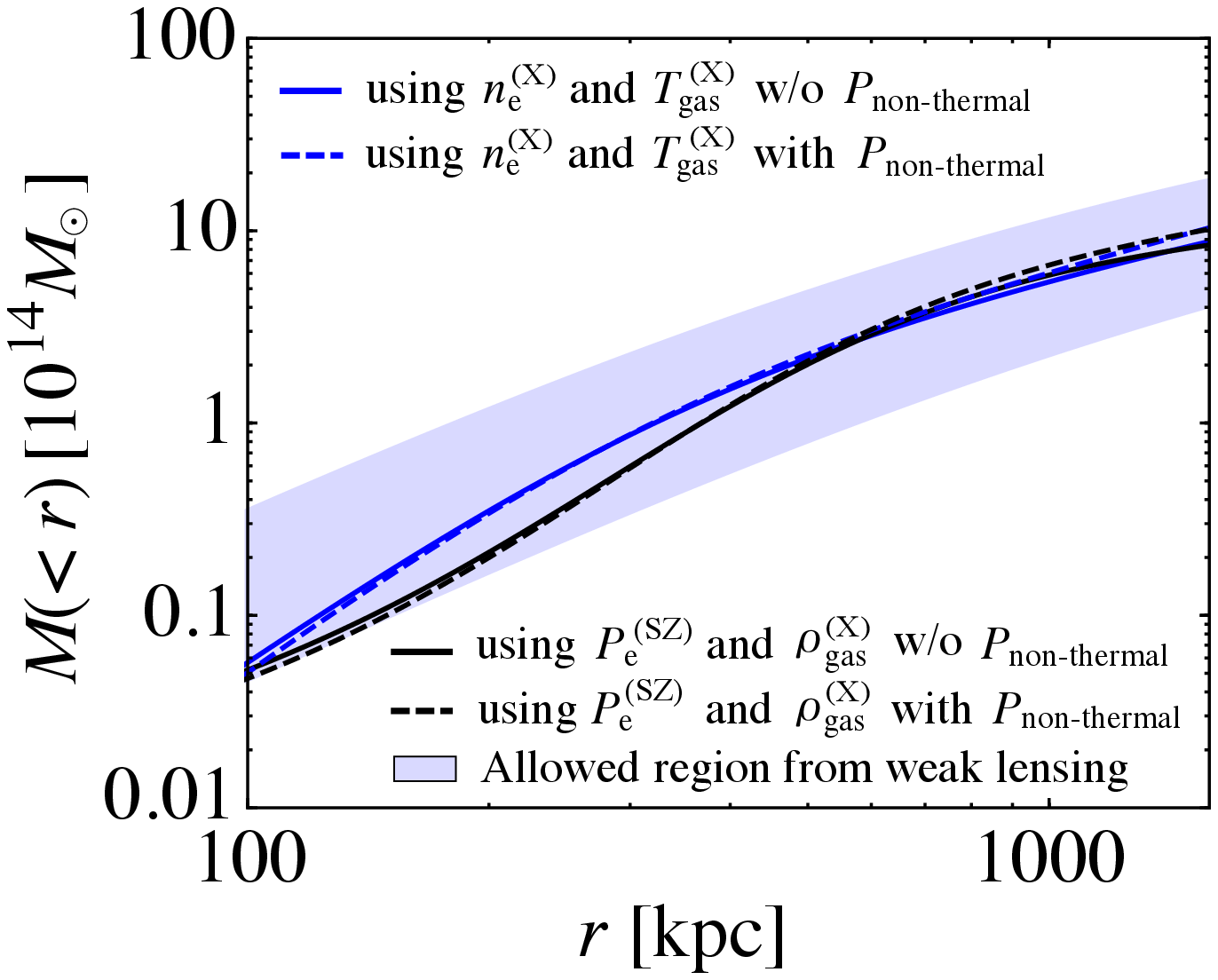}
\caption{
Radial mass profile of the Coma cluster.
The shaded region is the observationally allowed 1-$\sigma$ region from the WL observations of~\cite{Coma_7}.
The blue solid curve is the thermal mass component $M_{\rm thermal}$ estimated from the X-ray observations only, and the  black solid curve is $M_{\rm thermal}$ estimated from the combination of X-ray and SZ observations.
The blue dashed and black dashed curves correspond to the same colour solid lines, however, now including a large non-thermal pressure contribution.
\label{fig:Coma_M}
}
\end{center}

\begin{center}
  \includegraphics[width=0.60\hsize,clip]{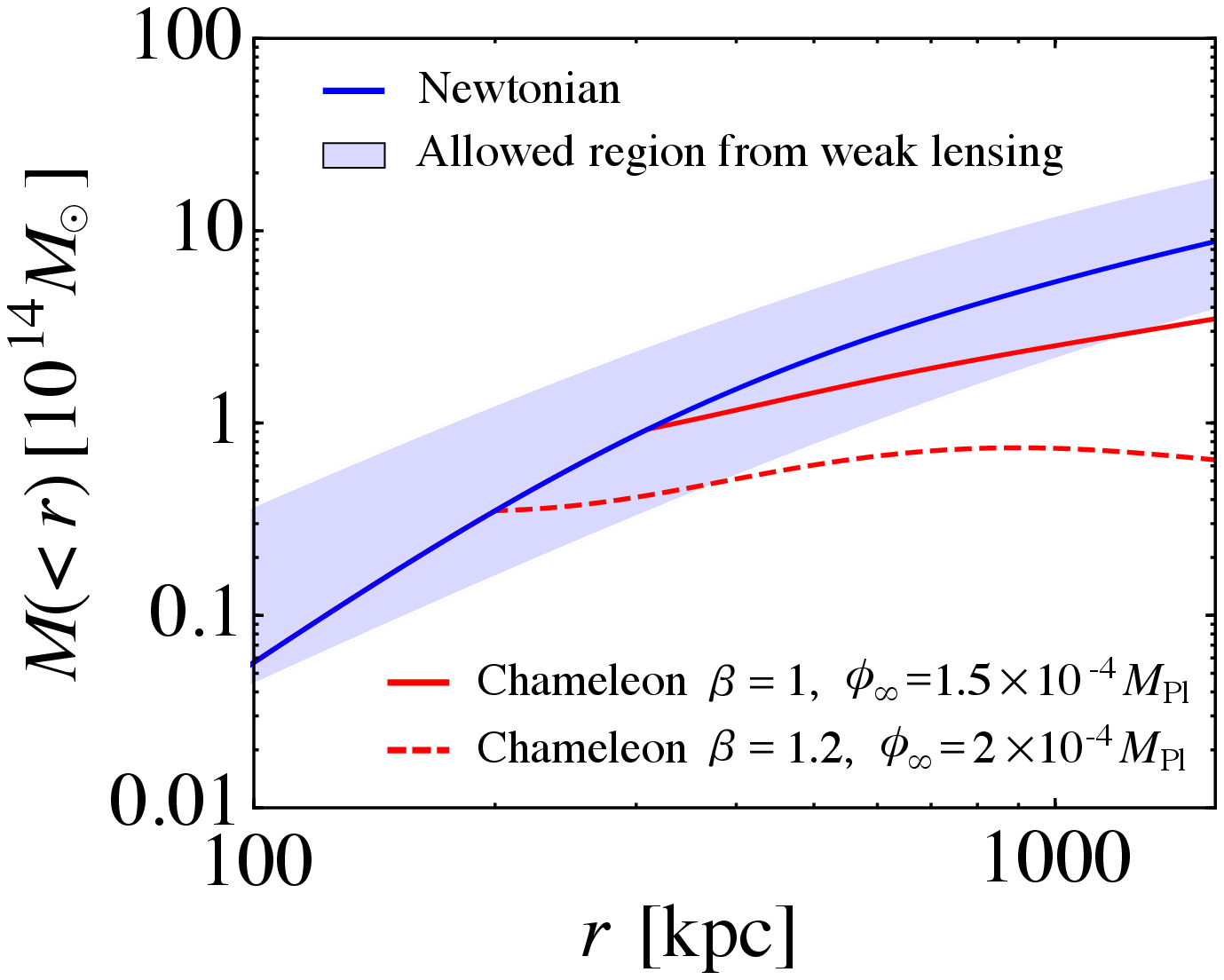}
\caption{
Same as Fig.~\ref{fig:Coma_M} but in the presence of the chameleon field.
The red solid and red dashed curves are the combination of the thermal mass and chameleon mass components, $M_{\rm thermal}+M_\phi$, when $(\beta,\phi_\infty/M_{\rm Pl})=(1,1.5\times 10^{-4})$ and $(1.2,2\times 10^{-4})$, respectively.
}
\label{fig:Coma_M2}
\end{center}
\end{figure}

Recently, Fusco-Femiano, Lapi, and Cavaliere \cite{Coma_1} analysed the consistency between the observations of the X-ray surface brightness, X-ray temperature, and SZ observations, adopting a ``Supermodel''.
The Supermodel yields a direct link between the X-ray and the SZ observations based on the entropy profile.
They report a tension between the pressure from the X-ray observations and that from SZ observation in the Coma cluster.
The authors argue that an additional non-thermal pressure resolves the tension.
In this paper, we adopt a similar observational data set and reconstruct the 3-dimensional gas profiles using the relations described in Appendix~\ref{sec:3dprofiles}.
We find a similar tension in our results and model a non-thermal pressure component as described in Sec.~\ref{sec:hydrostaticmass}, which however, is slightly different from the non-thermal pressure discussed in~\cite{Coma_1}.
The non-thermal pressure in~\cite{Coma_1} is a constant, which is independent of the radius.
The non-thermal pressure we introduce in Sec.~\ref{sec:hydrostaticmass} is a function of
radius, and its fraction in the total pressure becomes large only in the outer region.
Nevertheless, our models fit the data reasonably well and can be used to put a useful constraint
on the chameleon modification.
This is because we only use the limited data in the range of radii of $100~{\rm kpc}<r_\perp<1~{\rm Mpc}$, where the shape of the mass profile drives the constraints.

Fig.~\ref{fig:Coma_M} shows the different radial mass profiles reconstructed from the different gas observations and the lensing mass in Newtonian gravity, including effects from the non-thermal pressure introduced in Sec.~\ref{sec:hydrostaticmass}.
The blue solid curve is the hydrostatic mass from Eq.~(\ref{mrrt}) with the reconstructed $n_{\rm e}$$(=\rho_{\rm gas}(2+\mu)/5\mu m_{\rm p})$ and $T_{\rm gas}$ from the X-ray observations.
The black solid curve is the hydrostatic mass from Eq.~(\ref{mrrp}) with the reconstructed $\rho_{\rm gas}$ and $P_{\rm gas}$ from the X-ray observations and the SZ observations.
Finally, the shaded region in Fig.~\ref{fig:Coma_M} shows the allowed $1\sigma$-region of the WL mass profile fitted using a NFW density profile with $M_{\rm vir}=8.92^{+20.05}_{-5.17}$$\times10^{14}h^{-1} M_{\odot}$ and $c=3.5^{+2.57}_{-1.79}$.
At the scales of $100~{\rm kpc}<r<1~{\rm Mpc}$, the blue and black curves are consistent within the shaded region, while for $r<100$~kpc, the curves are out of the shaded region.
Thus, for $100~{\rm kpc}<r<1~{\rm Mpc}$, although the mass estimates differ up to the $50\%$ level, within the observational error of the lensing mass, the mass profiles estimated by the gas observations are consistent with each other and the lensing mass profile.
This suggests that  hydrostatic equilibrium is a good approximation for the outer region of the Coma cluster, given the error of the lensing measurement.
The discrepancies in the inner region $r<100~{\rm kpc}$ are a known problem in the mass reconstruction and beyond the scope of the present paper:
the validity of  hydrostatic equilibrium in the inner region has been investigated by many authors (see, e.g.,~\cite{Fabian1,Fabian2,FujitaOhira} and
references therein) with no consensus found.
Note, however, that the WL observations are not sensitive to the density profile in the inner region~\cite{Coma_7}.
We, therefore, base our analysis on a simple extrapolation of the NFW profile.
Recent lensing observations of the Coma cluster~\cite{Okabe2013} support the validity of this assumption for $100~{\rm kpc}< r < 1~{\rm Mpc}$ as well as
indicate its limitation for  $r<100~{\rm kpc}$.

In order to estimate the influence of the non-thermal pressure on the mass profile, the blue and black dashed curves in Fig.~\ref{fig:Coma_M} show the sum of the thermal mass profile $M_{\rm thermal}$ and the non-thermal mass component $M_{\rm non-thermal}$ determined by Eq.~(\ref{grg}).
The blue dashed curve is obtained from the X-ray observations via Eq.~(\ref{nonthmX}), whereas the black dashed curve is obtained from the combination of the SZ and X-ray observations from Eq.~(\ref{nonthmSZ}).
At $r=1$~Mpc, the non-thermal pressure enhances the total hydrodynamical mass estimation by a few $\times10\%$.
This reflects the limited effect of the non-thermal pressure predicted by hydrodynamical simulations.

Finally, we include the chameleon field in our mass comparison.
In Fig.~\ref{fig:Coma_M2}, we show the thermal radial mass profile and the combination with the chameleon mass component $M_{\rm thermal}+M_\phi$ (red curves).
The red solid and red dashed curve is obtained for
$(\beta,\phi_\infty/M_{\rm Pl})=(1,1.5\times 10^{-4})$ and for $(1.2,2\times 10^{-4})$, respectively.
These two sets of parameters of the chameleon model illustrate typical scenarios where the chameleon force causes a possible discrepancy between the gas and the lensing masses.
Note that these curves are determined from $M_{\rm thermal}$ and $M_\phi$ in Eq.~(\ref{HE11}), where $M_{\rm thermal}$ is reconstructed from the observational data and $M_\phi$ is given by Eq.~(\ref{chameleonmass}), and, therefore, the slightly oscillatory feature of the $\beta=1.2$ curve does not reflect any physically meaningful effect.
The blue curve represents the case without the chameleon force, which is close to the red solid curve and the red dashed curve in the inner region, where the chameleon field is suppressed.
Further out, the chameleon force reduces the hydrostatic mass $M_{\rm thermal}+M_\phi$ with respect to the mass obtained in Newtonian gravity because the chameleon force introduces an extra attractive force.
As is clear from this figure, we can only put a constraint on the chameleon model that influences
the gas distribution in the range $r\lesssim1$~Mpc.
The critical radius at which the chameleon force begins to contribute is determined by $\beta/\phi_\infty$ [see Eq.~(\ref{cp})] and the amplitude of the chameleon force is determined by $\beta$ (see Sec.~\ref{sec:chameleon}).
Thus, these two parameters in the chameleon models are constrained by comparing the hydrostatic mass and lensing mass under the assumption of hydrostatic equilibrium.

\subsubsection{Non-spherical symmetry} \label{sec:systematics2}

Next, let us consider systematic effects that can be introduced by deviations from spherical symmetry.
Here, we assume that the three dimensional profile of the electron number density,
temperature, and pressure are written as,
\begin{eqnarray}
&&n_{\rm e}(r,\theta,\varphi)=\bar n_{\rm e}(r)[1+\delta_{n_{\rm e}}(r,\theta,\varphi)],\\
&&T_{\rm gas}(r,\theta,\varphi)=\bar T_{\rm gas}(r)[1+\delta_{T_{\rm gas}}(r,\theta,\varphi)],\\
&&P_{\rm e}(r,\theta,\varphi)=\bar P_{\rm e}(r)[1+\delta_{P_{\rm e}}(r,\theta,\varphi)],
\end{eqnarray}
where $\delta_{n_{\rm e}} , \delta_{T_{\rm gas}}$ and $\delta_{P_{\rm e}}$ describe
deviation from the spherical symmetric profiles, $\bar n_{\rm e}(r)$,
$\bar T_{\rm gas}(r)$, and $\bar P_{\rm e}(r)$, respectively.

The effect of the clumpiness on the electron number density
can then be estimated as follows. Introducing an average
over the spherical symmetric profiles, we assume
$\langle \delta_{n_{\rm e}} \rangle=0$ and $\langle \delta_{n_{\rm e}}^2 \rangle \neq 0$.
Assuming that the temperature perturbation is negligible, i.e.,
$\delta_{T_{\rm gas}}=0$, the observed X-ray temperature profile is not changed.
The SZ profile is not affected by clumping either because $\langle \delta_{P_{\rm e}} \rangle
=\langle \delta_{n_{\rm e}} \rangle=0$ from the equation of state.
However, the surface brightness is increased by the clumpiness and can be rewritten as
\begin{eqnarray}
S_{\rm X} \propto \int n_{\rm e}^2dz = (1+\langle \delta_{n_{\rm e}}^2\rangle)\int \bar n_{\rm e}^2dz,
\end{eqnarray}
where $1+\langle \delta_{n_{\rm e}}^2 \rangle$ is referred to as the clumping factor.
This affects the reconstruction of the electron number density.
When the clumping factor is non-zero, $\bar n_{\rm e}$ is replaced by
$\bar n_{\rm e}/\sqrt{1+\langle \delta_{n_{\rm e}}^2 \rangle}$. Then, the thermal mass profile reconstructed
from observations of the SZ effect and X-ray surface brightness, Eq. (\ref{thermalSZ}), is
enhanced by a factor $\sqrt{1+\langle \delta_{n_{\rm e}}^2 \rangle}$.
However, the thermal mass profile reconstructed from X-ray observations, Eq. (\ref{thermalX}),
is not affected by the clumpiness.
In the case where $1+\langle \delta_{n_{\rm e}}^2\rangle  =1.5$, which corresponds to the estimated clumping factor for the A1835 cluster~\cite{clumpiness}, we have an
enhancement of the hydrostatic mass by a factor $\sim1.2$.
Thus, systematics from the clumpiness could be of order a few $\times 10 \%$.

Besides the clumpiness, large-scale spherical asymmetries of a cluster may
cause an additional systematic bias. Three dimensional ellipticity as well as
substructures of the Coma cluster have been studied in Ref.~\cite{Andrade}.
They have reported an ellipticity of the electron density in the Coma cluster
of $\epsilon=\sqrt{1-e^2}=0.84$ with eccentricity $e$
such that we can ignore the effect in our analysis.
Nonetheless the assumption of spherical symmetry introduces systematics errors,
which should be investigated in more quantitative detail in a future work.

%

%
\section{Summary and Conclusions} \label{sec:conclusion}

We have proposed a novel method to test gravity in the outskirts of galaxy clusters by comparing their hydrostatic and lensing mass estimates.
The hydrostatic mass profile of a cluster can be inferred from the 3-dimensional gas temperature, electron number density, and electron pressure profiles from the projected observations of the X-ray surface brightness, the X-ray temperature, and the SZ CMB temperature profile, by implementing a parametric reconstruction method.
The dark matter density profile can furthermore be constrained by WL observations.
Here, we adopt the NFW density profile to describe the dark matter distribution within the cluster.
In the case of hydrostatic equilibrium of the gas and standard gravity, the different mass estimates should agree.
In the presence of a chameleon field, coupling to the matter fields and introducing an attractive fifth force, the relation between the mass estimated from the gas observations and from lensing changes, and can therefore be used as a test of gravity.

Combining measurements of the X-ray surface brightness, the X-ray temperature, the SZ effect, and lensing of the Coma cluster, we performed an MCMC analysis of the model parameter space, describing the cluster profiles and gravity theory, and have obtained competitive constraints on the chameleon gravity model parameters $\beta$ and $\phi_\infty$, the coupling strength of the chameleon field and the field value in the environment of the cluster, which we approximate here by the cosmological background.
Contrary to a previous study in~\cite{Terukina} that constrains the modified gas distribution in the Hydra A cluster measured through the X-ray temperature, our new constraint does not rely on the assumption of a polytropic equation of state of the gas, employs a Bayesian statistical approach for inferring parameter constraints on the full set of model parameters, and yields a tighter bound on the modified gravity parameters than these previous results through the combination of the X-ray, SZ, and lensing observations available for the Coma cluster.
We emphasise that our results provide a powerful constraint on $f(R)$ gravity models, corresponding to a particular choice of the chameleon coupling constant $\beta=\sqrt{1/6}$, for which we obtain an upper bound of $|f_{\rm R0}|\simlt 6\times 10^{-5}$ at the 95\% CL.
This bound is competitive with the current strongest cosmological constraints inferred on $f(R)$ gravity (see Fig.~\ref{fig:mg_limit}).

An important systematic that can affect our analysis can be introduced by deviations from hydrostatic equilibrium of the cluster gas.
We have therefore carefully examined the validity of hydrostatic equilibrium in the Coma cluster.
Assuming Newtonian gravity, we compare the different mass estimates from the three different gas observations and the WL mass.
We find that the mass profiles from the gas and WL observations can deviate from each other by up to 50\% but that they are consistent within the observational errors of the lensing measurement.
We analyse the effect of including a non-thermal pressure component, with a radial profile calibrated to hydrodynamic simulations but with extremised amplitude.
This contribution only marginally affects our reconstructed masses, and we conclude that  hydrostatic equilibrium is a good approximation to describe the outer region of the Coma cluster.
Note, however, that the effect from the chameleon force on the hydrostatic mass is opposite to the effect of the non-thermal pressure.
Hence, the chameleon force can compensate for a large contribution from non-thermal pressure and cause a degeneracy between the two effects.
On the other hand, the magnitude of the non-thermal pressure that would be required to compensate for the effects of the chameleon force tested here is not to be expected from current hydrodynamical simulations.
It is, however, not clear whether the presence of a chameleon field could significantly enhance the non-thermal pressure contribution in the Coma cluster such that it could cancel the effects of the chameleon field, and act to alleviate the constraints on the modification of gravity.
In this regard, it will be useful to analyse the non-thermal pressure of chameleon gravity models using hydrodynamical simulations along with a more detailed study of the Newtonian case.
As for $f(R)$ gravity, such hydrodynamical simulations have recently been conducted by Arnold \emph{et al.}~\cite{Arnold}.
They estimate the non-thermal pressure from the bulk motion in the intracluster medium and find that it only leads to substantial contributions in merging clusters, which can be identified and excluded to obtain statistical quantities like X-ray and SZ scaling relations.
Their results suggest that the effects of non-thermal pressure in a relaxed cluster like Coma are not significant, at least in the case of the $f(R)$ gravity models.

Further effects which may cause deviations from the hydrostatic equilibrium have been discussed in~\cite{Coma_1, clumpiness, DSuto}.
Ref.~\cite{DSuto} found that the mass estimated under the hydrostatic equilibrium assumption deviates from the true mass on average
by $\sim (10 - 20)\%$ fractionally in a simulated halo due to gas acceleration.
Given the large errors on the measurement of the lensing mass of the Coma cluster, we can ignore this deviation in our current analysis.
Future measurements such as from the Astro-H X-ray observations will allow more precise modelling of the Coma cluster.

Our results demonstrate that galaxy clusters are useful probes of gravity.
The method described in this paper may be applied to other clusters.
However, one should be cautious about the individual properties of each cluster;
the assumptions adopted in the present paper might not be guaranteed for other galaxy
clusters and need to be considered for each case.
The key is to understand the motion and distribution of the gas component in clusters;
the combination of multi-wavelength observations, as demonstrated by the recent results
by the \emph{Planck} satellite~\cite{Plancka, Planck_cl,Planck_cl2,Planck2},
will provide a clue on how to solve this difficult issue.
In the near future, we will have stacked lensing, SZ, and X-ray profiles for hundreds of
clusters. The combination of multi-wavelength observations for many clusters will significantly improve the tests of gravitational interactions on cluster scales.

\vspace{2mm}
We thank Y.~Fukazawa, Y.~Suto, D.~Suto, T.~Kitayama, T. Kobayashi for support and useful discussions. We thank A. Vikhlinin and E. Churazov for providing the X-ray data used in this paper. LL and KK were supported by the European Research Council starting grant. LL acknowledges support from the STFC Consolidated Grant for Astronomy and Astrophysics at the University of Edinburgh. The research by KY is supported in part by Grant-in-Aid for Scientific researcher of Japanese Ministry of Education, Culture, Sports, Science and Technology (No.~21540270 and No.~21244033). DB, KK and BN are supported by STFC grants ST/K00090/1 and
KK is also supported by STFC grant ST/L005573/1. KK thanks the Leverhulme trust for its support. KY thanks Prof. L. Amendola and the people at ITP of Heidelberg University for their hospitality during his stay. This work is supported by exchange visits between JSPS and DFG.

\begin{appendix}
\section{Reconstruction of the 3-dimensional gas profiles}
\label{sec:3dprofiles}
We summarise the reconstruction method for the 3-dimensional profiles
of the gas density, temperature, and pressure, using observations of the X-ray
temperature, surface brightness, and SZ effect used in Sec.~\ref{sec:mcmc} to derive constraints on chameleon gravity from the Coma cluster.
We begin with a short discussion on the quantities that are observed in the X-ray
measurements and through the SZ effect.

{}From X-ray observations, one obtains the projected X-ray temperature profile
\begin{eqnarray}\label{xt}
  T_{\rm X}(r_\perp)&=&\frac{\int W\left(\sqrt{r^2_\perp+z^2}\right)
  T_{\rm gas}\left(\sqrt{r_\perp^2+z^2}\right)dz}
  {\int W \left(\sqrt{r_\perp^2+z^2}\right)dz},
\end{eqnarray}
where $r_\perp$ is the radius perpendicular to the line of sight direction, and $W(r)$ is the weight factor,
which may be written as $W(r)=n_{\rm e}^2(r)T_{\rm gas}^{1/2}(r)$ for the emission weighted temperature and as
$W(r)=n_{\rm e}^2(r)T_{\rm gas}^{-3/4}(r)$ for the spectroscopic-like temperature \cite{Mazzotta}.
Here, $n_{\rm e}(r)$ denotes the electron density profile.
In this paper, we use the emission weighted temperature, but we checked that the 3-dimensional temperature and electron number density profiles do not depend on the choice between the two weightings.

The X-ray surface brightness is given by
\begin{eqnarray}
\label{sb}
  S_{\rm X}(r_\perp)&=&\int n_{\rm e}^2\left(\sqrt{r_\perp^2+z^2}\right)
  \lambda_{\rm c}\left(T_{\rm gas}\right) dz,
\end{eqnarray}
where $\lambda_{\rm c}(T_{\rm gas})$ is the cooling function.
To estimate the cooling function, we used XSPEC \cite{xspec}
adopting the thermal plasma emission spectra model with the APEC code \cite{apec}.
The XSPEC software gives the X-ray flux based on the APEC model corresponding to
the observational band from $0.5 {\rm keV}$ to $2.5 {\rm keV}$ \cite{Coma_2}.
The X-ray flux can be converted to the cooling function by the flux-luminosity relation.
The metal abundance in the innermost region of the cluster is larger than in the outer region, $Z=0.4Z_\odot$~\cite{Coma_1} and $Z=0.3Z_\odot$~\cite{Coma_8}, respectively.
However, as the difference is small and does not affect our constraints, we adopt a metal abundance of $Z=0.3Z_\odot$ throughout the cluster.

Photons from the CMB passing through clusters are scattered by the hot gas,  and this distorts the CMB spectrum as a function of frequency.
This SZ effect yields a contribution to the CMB temperature of
\begin{eqnarray}\label{sz}
\Delta T_{\rm SZ}(r_\perp)&=&-2T_{\rm CMB}
\frac{\sigma_{\rm T}}{m_{\rm e}}\int P_{\rm e}
\left(\sqrt{r^2_\perp+z^2}\right) dz,
\end{eqnarray}
where $T_{\rm CMB}=2.725K$ is the CMB temperature, $\sigma_{\rm T}$ is the Thomson cross section, $m_{\rm e}$ is the electron mass, and $P_{\rm e}(r)$ is the electron pressure.

Having summarised the quantities observed in the X-ray and SZ measurements, we now use them to reconstruct the 3-dimensional gas density, temperature, and pressure profiles.
For this purpose, we adopt the following fitting functions for the 3-dimensional profiles of $T_{\rm gas}(r)$, $n_{\rm e}(r)$, and $P_{\rm e}(r)$.
For $T_{\rm gas}(r)$, we use the fitting formula calibrated to numerical simulations \cite{Burns:2010sx}
 \begin{eqnarray}\label{tf}
   &&T_{\rm gas}(r)=T_0\left[1+A\left(\frac{r}{r_0}\right)\right]^{b_0},
 \end{eqnarray}
where $T_0$, $A$, $r_0$, and $b_0$ are free parameters.
For the electron number density, we assume a simple isothermal $\beta$ model~\cite{Cavaliere:1976tx}
\begin{eqnarray}
   &&n_{\rm e}(r)
     =n_0\left[1+\left(\frac{r}{r_1}\right)^2\right]^{b_1},
   \label{nf}
 \end{eqnarray}
where the free parameters are $n_0$, $r_1$, and $b_1$.
Finally, we adopt the generalised NFW profile for the pressure proposed by~\cite{Nagai:2007mt},
 \begin{eqnarray}\label{pf}
   P_{\rm e}(r)=\frac{P_0}{(r/r_2)^{b_2}(1+(r/r_2)^{b_3})^{b_4}},
 \end{eqnarray}
for the 3-dimensional electron pressure profile with the fitting parameters $P_0$, $r_2$, $b_3$, and $b_4$.

We compute the projected profiles in Eqs.~(\ref{xt}), (\ref{sb}), and (\ref{sz}) with the
fitting functions of Eqs.~(\ref{tf}), (\ref{nf}), and (\ref{pf}), and determine the best fit parameters
$T_0$, $A$, $r_0$, $b_0$, $n_0$, $r_1$, $b_1$, $P_0$, $r_2$, $b_2$, $b_3$ and $b_4$ by comparing the profiles with the observations from the X-ray temperature, X-ray surface brightness, and SZ effect of the Coma cluster in Sec.~\ref{sec:systematics1}. In this way, we obtain the reconstructed 3-dimensional gas density, temperature, and pressure profiles of the cluster.

Note that since we assume the hydrostatic equilibrium Eq.~(\ref{hydroconstraints}) in the MCMC analysis in Sec.~\ref{sec:mcmc}, we only need to define two of these profiles, of which one can also be the matter density profile and from which the other profiles can be derived, however, not necessarily reproducing the exact analytic expressions of the fitting functions.
In Sec.~\ref{sec:mcmc}, we choose to work with the electron number density Eq.~(\ref{nf}) and the NFW profile Eq.~(\ref{nfwfit}).
The choice of the NFW profile simplifes the computation of the chameleon force and allows the use of the analytic approximation derived in Sec.~\ref{sec:chameleon}.
Hence, the degrees of freedom reduce to $T_0$, $n_0$, $r_1$, $b_1$, including the NFW parameters $M_{\rm vir}$ and $c$ as well as the chameleon model parameters $\beta$ and $\phi_{\infty}$ (or $\beta_2$ and $\phi_{\infty,2}$), where $T_0$ is required to set the integration constant in Eq.~(\ref{pthermal}).
This approach yields reasonable reduced $\chi^2$ values when fitted to the observational data in Sec.~\ref{mcmcanalysis}.

%

\begin{figure}[t]
\begin{center}
\includegraphics[angle=-90,width=0.7\hsize,viewport= 0 0 1650 900,clip]{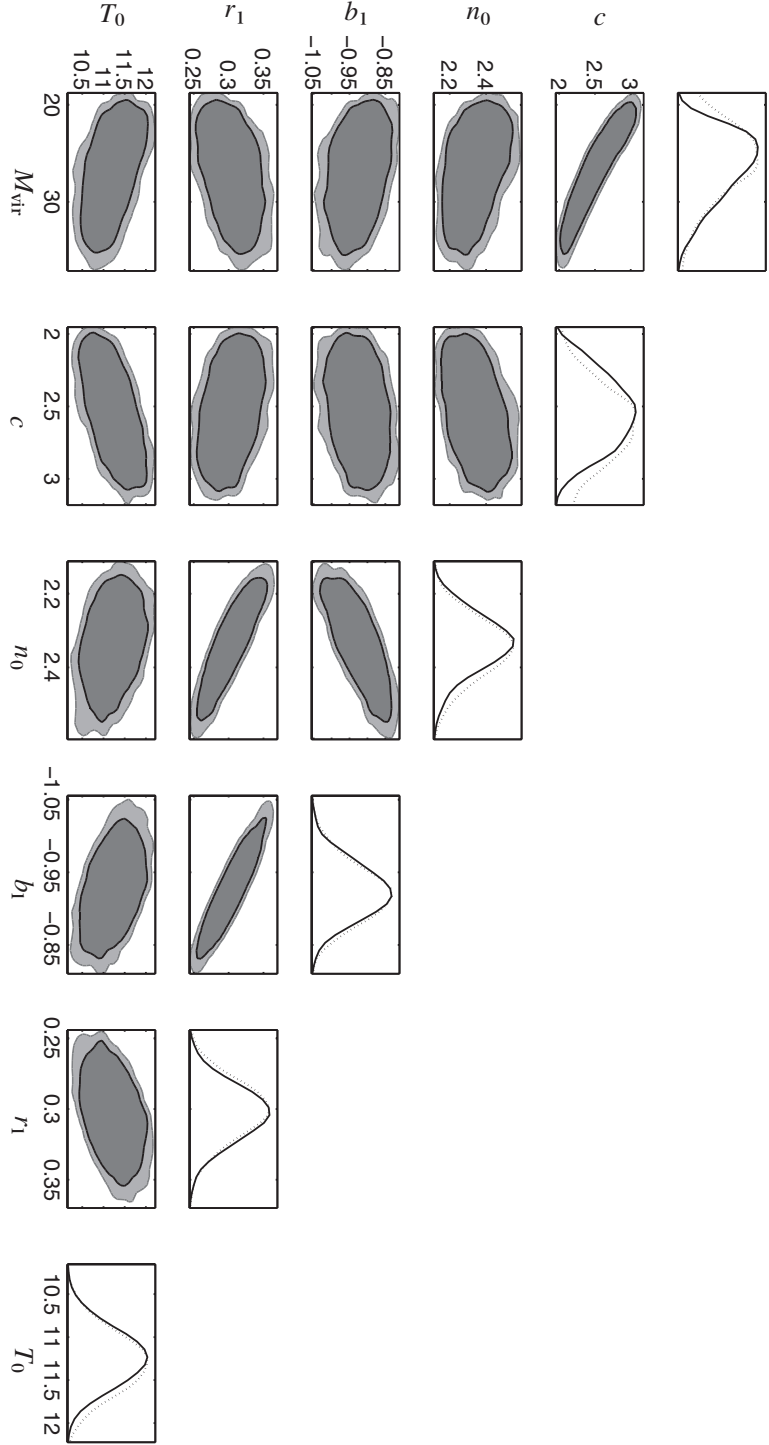}
\caption{
95\% (deep gray region) and 99\% CL (pale gray region) 2-dimensional marginalised contours of the $6$ model parameters
$T_0~{\rm [keV]}$, $n_0~{\rm [10^{-2}cm^{-3}]}$, $b_1$, $r_1~{\rm [Mpc]}$, $M_{\rm vir}~{[10^{14} M_\odot]}$,
and $c$ in the Newtonian scenario, obtained from the MCMC analysis, using the joint set of X-ray, SZ, and WL data.
The most-right panels of each row show the 1-dimensional marginalised constraints (solid) and likelihood distributions (dotted).
\label{fig:mcmc1x}
}
\end{center}
\end{figure}

\begin{figure}[t]
\begin{center}
\includegraphics[angle=-90,width=0.75\hsize,viewport= 0 0 1650 900,clip]{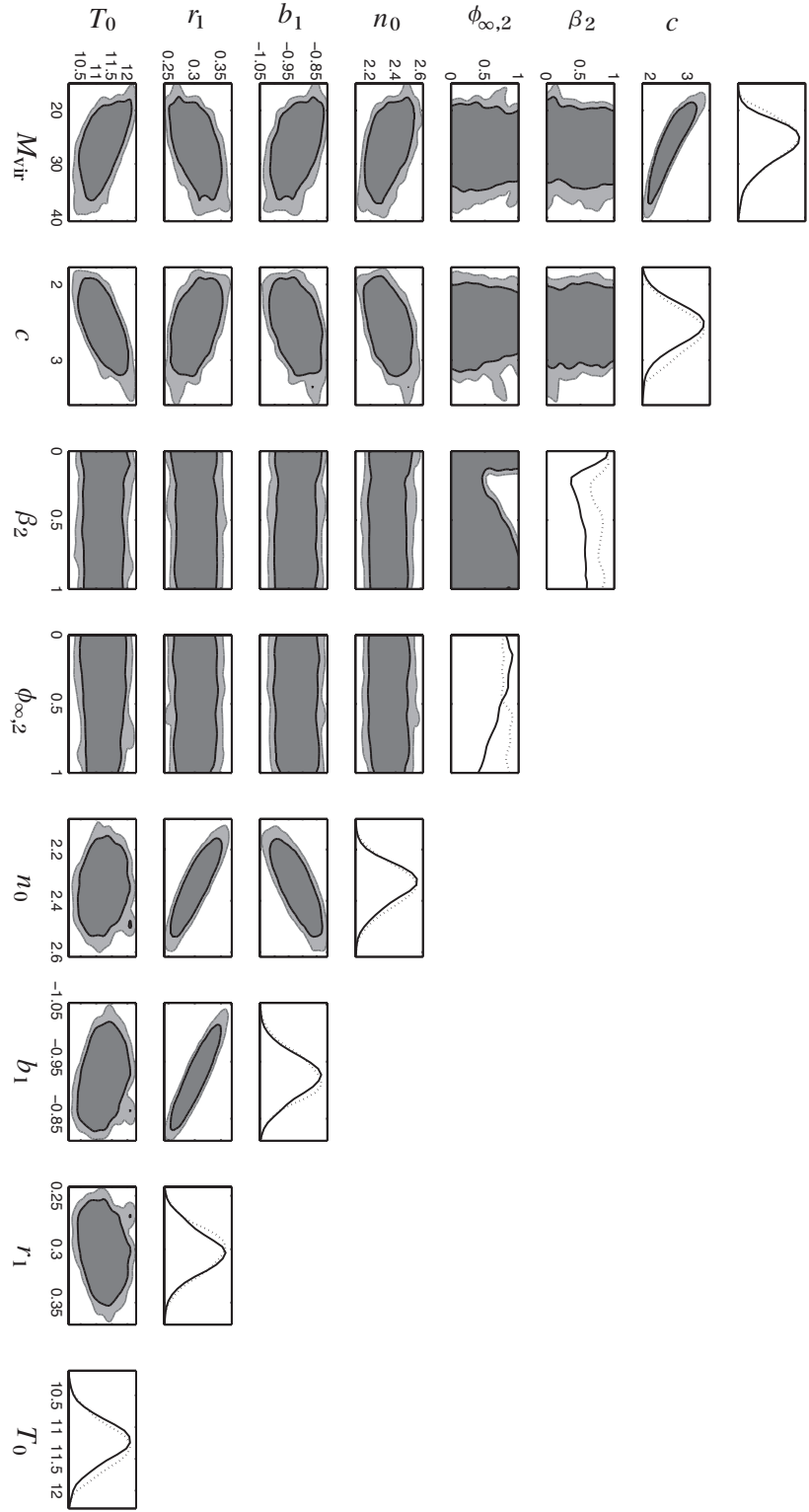}
\caption{
Same as Fig.~\ref{fig:mcmc1x} but when including the chameleon parameters $\beta_2$ and $\phi_{\infty,2}$ in the MCMC analysis.
The 2-dimensional marginalised contours of $\beta_2$ and $\phi_{\infty,2}$ are also shown in Fig.~\ref{fig:Coma_phinf-beta2x}.
\label{fig:mcmcx}}
\end{center}
\end{figure}


\end{appendix}

\end{document}